\documentclass[twocolumn,iop,numberedappendix]{emulateapj}
\newcommand{\isIncluded}{true}
\usepackage{amsmath,amssymb}
\usepackage{listings}
\usepackage{graphicx}
\usepackage{multirow}
\usepackage{booktabs,tabu}
\usepackage{natbib}
\usepackage{hyperref}
\usepackage[x11names]{xcolor}

\newcommand{\mmin}{\ensuremath{m_\text{min}}}
\newcommand{\Hs}{\ensuremath{\mathcal{H}_\star}}
\newcommand{\hs}{\ensuremath{h_\star}}

\newcommand{\Arc}{\ensuremath{A_{\rho_c}}}
\newcommand{\thp}{\ensuremath{\vec{\theta}}}
\newcommand{\psp}{\ensuremath{\vec{\psi}}}

\newcommand{\emcee}{\textsc{emcee}}

\newcommand{\ms}{\ensuremath{m_\star}}
\newcommand{\Ms}{\ensuremath{\mathcal{M}_\star}}

\shorttitle{Empirical HMF}
\shortauthors{Murray et al.}

\begin{document}
\newcommand{\mypref}{}
	
	
	\title{An Empirical Mass Function Distribution}
	
	
	\author{S.~G.~Murray\altaffilmark{1,2,3}, A.~S.~G.~Robotham\altaffilmark{1} and C.~Power\altaffilmark{1,2}}
    
%
%
	
	
	\altaffiltext{1}{International Centre for Radio Astronomy Research (ICRAR), University of Western Australia, Crawley, WA, 6009, Australia}
	\altaffiltext{2}{ARC Centre of Excellence for All-sky Astrophysics}
	\altaffiltext{3}{International Centre for Radio Astronomy Research (ICRAR), Curtin University,  Bentley, WA 6102, Australia}
		
	
%






\begin{abstract}

The halo mass function, encoding the comoving number density of dark matter halos of a given mass, plays a key role in understanding the formation and evolution of galaxies. As such, it is a key goal of current and future deep optical surveys to constrain the mass function down to mass scales which typically host $L_\star$ galaxies. Motivated by the proven accuracy of Press-Schechter-type mass functions, we introduce a related but purely empirical form consistent with standard formulae to better than 4\% in the medium-mass regime, $10^{10}-10^{13} h^{-1}M_\odot$. In particular, our form consists of 4 parameters, each of which has a simple interpretation, and can be directly related to parameters of the galaxy distribution, such as $L_\star$.

Using this form within a hierarchical Bayesian likelihood model, we show how individual mass-measurement errors can be successfully included in a typical analysis whilst accounting for Eddington bias. 
We apply our form to a question of survey design in the context of a semi-realistic data model, illustrating how it can be used to obtain optimal balance between survey depth and angular coverage for constraints on mass function parameters.

Open-source {\tt Python} and {\tt R} codes to apply our new form are provided at \url{http://mrpy.readthedocs.org} and \url{https://cran.r-project.org/web/packages/tggd/index.html} respectively. 
\end{abstract}

\section{Introduction}
\label{mrp:sec:intro}

Our understanding of galaxy evolution is largely hinged upon the galaxies' dark scaffolding -- the dark matter distribution and its network of filaments, halos and voids. In particular, every galaxy is believed to be housed in a dark matter halo, each the product of a turbulent series of mergers over cosmic time. This history of mergers gives rise to an elegant picture of cosmic structure formation: the so-called hierarchical paradigm \citep{White1978}. This paradigm has been helpful in allowing predictions of many key characteristics of the halo distribution, including the halo mass function (HMF): the comoving number density of haloes as a function of their mass.

Clearly, the evolution of the halo distribution and that of the hosted galaxies is fundamentally interconnected. For this reason, measurement of the HMF and its connection to observed galaxy properties are of fundamental importance. Whilst the abundance of cluster-mass halos is tightly related to the underlying cosmology \citep[eg.][]{Bocquet2016,DeHaan2016}, the galaxy population is predominantly hosted by haloes in the medium-mass regime $\sim 10^{10} - 10^{13}\ h^{-1}M_\odot$. Measurement of the HMF in this regime via dynamical mass estimates is thus a key goal for current surveys such as GAMA \cite[Galaxy and Mass Assembly, ][]{Driver2011} and future deep surveys such as WAVES \cite[Wide Area VISTA Extra-galactic Survey, ][]{Driver2016} and MSE \cite[Manukea Spectroscopic Explorer, ][]{McConnachie2016}.

The utility of measuring the HMF with dynamical halo masses inferred from groups of galaxies has been shown in \cite{Eke2006}, who used the 2PIGG (2dFGRS Percolation-Inferred Galaxy Group) catalogue. Using a simple empirical correction to mitigate Eddington bias, they were able to infer the group luminosity function and mean mass-to-light ratio of galaxies within their sample. Additionally, they were able to infer a best-fit value of $\sigma_8$ by comparing to simulation results. However, similar studies with improved analysis methods and larger surveys have not been forthcoming. Instead, there has been an increased focus on less direct methods of connecting the dark sector to galaxies, such as the halo occupation distribution \citep[HOD;][]{Berlind2003,Zehavi2005,Zehavi2011}. One reason for this focus may be an increased reliance on the standard formalism for theoretical HMFs.

The standard model for the HMF arises from physical arguments proposed by \cite{Press1974}, and we shall refer to this framework (with its many specific fits) collectively as the ``PS formalism" (not to be confused with the specific original form of \citet{Press1974}) and we shall refer to any fit within this framework as a ``PS-type" fit. In this model, the HMF can be written
\begin{equation}
	\frac{dn}{dM} = \frac{\rho_0}{M^2}\left|\frac{d \ln \sigma}{d\ln M}\right| f(\sigma),
\end{equation}
where $\rho_0$ is the universal mean matter density, and the model is dependent on the underlying cosmology via the mass variance, $\sigma^2(m)$, which is the second moment of the matter power spectrum, smoothed (by an arbitrary filter $W$) on a physical scale $R$ corresponding to $m$:
\begin{equation}
	\label{eq:sigma}
	\sigma^2(R) = \frac{1}{2\pi^2} \int_0^\infty dk\ k^2 P(k) W^2(kR).
\end{equation}
Assuming spheroidal collapse of halos, \cite{Press1974} derived a form for $f(\sigma)$ given by
\begin{equation}
	f(\sigma) = \sqrt{\frac{2}{\pi}}\frac{\delta_c}{\sigma}\exp\left(-\frac{\delta_c^2}{2\sigma^2}\right),
\end{equation}
where $\delta_c\approx1.686$ is the critical overdensity for spherical collapse. 

The PS formalism has proven remarkably successful in describing the HMF measured in cosmological simulations, albeit with modifications of $f(\sigma)$ to better capture the non-spherical nature of halo collapse \citep[eg.][]{Sheth2001}. However, though it appears to be universal with respect to cosmology and redshift, it has been shown to exhibit non-universality at the $\sim20$\% level, especially for halos defined by a spherical-overdensity criterion \citep{Tinker2008,Watson2013}. Furthermore, the best-fit seems to depend not only on underlying physics and halo definition, but on the algorithm used for locating halos within cosmological simulations \citep{Knebe2011}. Though these uncertainties do not prevent the use of the PS formalism in accurate fits to cluster abundance \citep{DeHaan2016}, they motivate the use of a simpler empirical distribution in the context of observed HMFs at lower masses.

In addition to this motivation, the most common descriptions of ensemble galaxy properties are entirely empirical; both luminosity functions (LFs) and galaxy stellar mass functions (GSMFs) are parameterised by single- or double-Schechter functions \cite{Schechter1976}, which capture their forms adequately without any first-principles motivation. Thus for example, in the single-Schechter case, the luminosity function is commonly expressed as
\begin{equation}
	\frac{dn}{dL} = \phi_\star \left(\frac{L}{L_\star}\right)^{-\alpha}\exp\left(-\frac{L}{L_\star}\right).
\end{equation}
Here the model parameters themselves hold explanatory power: $L_\star$ is a characteristic turn-over luminosity, and $\alpha$ is a power-law slope for low-luminosity galaxies. Measuring the HMF via the PS formalism obfuscates the relation between parameters and data, rendering the comparison of the dark HMF with the visible galaxy properties less clear. Given the apparently similar forms of the HMF, LF and GSMF, it would seem beneficial to construct their parameterisations in a similar fashion so as to facilitate comparison; for example, can we relate the evolution of a typical halo mass $\Hs$ with that of the typical stellar mass, $M_\star$, or luminosity $L_\star$?

Using a simple empirical form defined directly in terms of mass also bears other advantages. 
It is portable, both in the sense that anyone can easily create their own numerical function for its evaluation, and that data described by the form are reduced to a small set of interpretable parameters. 
It is fast, since it is a direct calculation. And since it is analytic, it affords more extended analyses and methods.

Conversely, the empirical nature of the form means that it is not easily linked with an underlying cosmology. For this reason, it will not be useful for cluster abundance studies which primarily seek to constrain the power spectrum normalisation $\sigma_8$, and dark energy parameters. Rather, it is most usefully applied to questions about the connection between dark halos and galaxies.

Indeed, such a form would seem to be appropriately used in empirical low-mass HMF studies in a similar way to how the Schechter function is used for luminosity function or galaxy mass function studies.
That is, the results of HMF analyses can simply be expressed as a collection of a small number of conceptually informative parameters with their uncertainties -- and furthermore these can be directly compared
to parameters of the galaxy population. 
In this way, the form would be seen as an observer's tool, rather than a theoretician's model.

In this paper, we propose an empirical form that is equivalent to a re-normalised generalised gamma distribution (hereafter GGD). The re-normalised GGD (which throughout we term the MRP, an acronym for the author's surnames) is a generalisation of the Schechter function, in which the sharpness of the exponential cut-off is variable. This similarity enables direct comparison between the HMF and galaxy ensemble properties (cf. \S\ref{mrp:sec:smhm}). 
Furthermore, its simple analytical nature offers a number of advantages in estimation of its parameters via hierarchical Bayesian models, which serve to account for potential distortions such as Eddington bias.

After introducing the form and its basic properties in \S\ref{mrp:sec:model}, we outline a fully general method of parameter estimation with the MRP in \S\ref{mrp:sec:fitting}, including arbitrary measurement uncertainties and selection effects. Using these methods, in \S\ref{mrp:sec:accuracy} we assess the accuracy of the MRP form with respect to both prediction of the PS formalism and large cosmological simulations.

Following this, we outline two simple ways in which the MRP may be linked to physical parameters: firstly in \S\ref{mrp:sec:dependence}, we provide general fitting functions which relate the MRP to the underlying cosmology. Following this in \S\ref{mrp:sec:smhm} we develop a generic form for the Stellar-Mass Halo-Mass relation, which connects halos with their stellar mass content, and which in the case of the MRP can be specified exactly in both mass limits. 

After this laying out of methodology and justification of our use of the MRP, in \S\ref{mrp:sec:survey} we set out to illustrate application of the MRP to questions of survey design, in the process utilising many of the results of the previous sections.
Finally, we conclude in \S\ref{mrp:sec:conc}.

To accompany this paper, we have developed an extensive Python code, called \textsc{mrpy}\footnote{Found at \url{http://mrpy.readthedocs.org}. Note that wherever possible, we provide the exact code to reproduce the figures in this paper as examples at \url{https://github.com/steven-murray/mrpy/examples}.}, which implements all of the functionality explored throughout the paper. Furthermore, we provide a reduced R version of this code which focuses on the basic statistical applications, called \textsc{tggd}\footnote{Found at \url{https://github.com/asgr/tggd} and downloadable from CRAN as \texttt{tggd}.}. 

\section{The MRP distribution}
\label{mrp:sec:model}
The MRP has four parameters, and can be written 
\begin{align}
\label{mrp:eq:g}
   \frac{dn}{d\log_{10} M} &\equiv \frac{dn}{dm} \equiv \phi(m|\vec{\theta}) \\ \nonumber
   &=  A \beta 10^{(\alpha+1)(m-\hs)} \exp\left(-10^{\beta(m-\hs)}\right) \\ \nonumber
   &\equiv g(m|\thp)
\end{align}
where $m = \log_{10} M$ is the logarithmic mass and $\thp$ is the parameter vector $(\hs\equiv\log_{10}\Hs,\alpha,\beta,\ln A)$. 
We purposely set $\phi$ as the generalised number density as a function of logarithmic mass, while for convenience we denote the specific MRP form as $g(m|\thp)$.

We note that $g$ is a generalisation of the Schechter function, which has a fixed $\beta=1$, so that many of the results derived in this paper will be applicable (in modified form) to the Schechter function also.

Usually, the GGD is defined for $A>0$, $\beta>0$, and $\alpha>-1$, where the latter two constraints ensure convergence at the high- and low-mass ends respectively. 
We dispense with the constraint on $\alpha$, under the assumption that a turnover occurs well below the observed mass limits to ensure convergence.
Indeed, typical HMFs exhibit $\alpha \sim -2$, and we note that for $\alpha \leq -2$ the GGD is divergent in total mass density if not lower-truncated.
Note that it is also common for HMFs defined under the PS formalism to be divergent in the lower limit \citep{Jenkins2001}, with the understanding that extra modelling is required below scales which we have currently probed in cosmological simulations.
 
\subsection{Relationship to PS formula}
\label{mrp:sec:rel_to_ps}
The MRP form can be precisely recovered via the PS formalism if we assume the analytic fit of \citet{Press1974} and that $\sigma(m)$ can be approximated by a power-law, $(m/m_1)^{-\gamma}$. Doing so yields the following equalities:
\begin{equation}
\left\{\Hs, \alpha,\beta, A\right\} = \left\{m_1 \sqrt{a}, \gamma-2, 2\gamma, \frac{a}{m_1 \sqrt{\pi}}\right\}
\end{equation}	
where $a = (2/\delta_c^2)^{1/\gamma}$. In practise, the MRP is not so restrictive -- $\alpha$ and $\beta$ are free -- so it is able to achieve more precise fits than this approximation would suggest.

We note also that the GGD naturally arises by considering the sum of exponential variates, raised to the power $\beta$. 
However, it is unclear how such a process relates to the HMF. 
Nevertheless, this composition of the GGD from simpler distributions motivates an expectation that a comprehensible chain of stochastic events leads to the HMF intrinsically having a form close to the GGD. 

\subsection{Cumulative Distribution}
An important quantity is the integral of $g$ over mass, which determines the normalisation of the distribution. 
Two cases are important here. When $\alpha>-1$, which is not typical of the HMF, but may be achieved in practice by a power-law selection function (see next section), we have
\begin{equation}
	\int g(m|\thp) dm =  A \Hs \Gamma(z_0) \equiv q_0(\thp).
\end{equation}
When $\alpha<-1$, the integral is non-convergent, and we present the result for a distribution truncated at some lower mass:
\begin{equation}
    \int_{\mmin}^\infty g(m|\thp)dm = A \Hs \Gamma(z_0,x) \equiv q(\thp,m_\text{min}),
\end{equation}
where 
\begin{align}
z_n &= (\alpha+1+n)/\beta, \\
x &= \left(\frac{\mathcal{M}_\text{min}}{\Hs}\right)^\beta \equiv 10^{\beta(m-\hs)},
\end{align}
and $\Gamma$ is the Euler (upper-incomplete) Gamma function. We will use $z_n$ and $x$ in this manner throughout the paper, noting that $x_m$ will refer to $x$ with $m_\text{min}$ replaced by arbitrary $m$, and $z$ without subscript refers to $z_0$.

We will typically deal with the latter case in this paper, as it is the more general of the two. 
It should be noted that for $z>0$, $\Gamma(z,0)=\Gamma(z)$, so that $q_0$ is merely the expected special case of $q$, under the obvious proviso that the integral converges.
Thus we need only specify results for the truncated case, and the special case can be read in.

We note that many numerical libraries only implement $\Gamma(z,x)$ for $z\geq0$, so that $\alpha$ is required to be $>-1$ (consistent with the usual definition of the GGD)\footnote{Notable exceptions are the GNU scientific library (GSL), which the R \textsc{tggd} package interfaces with, and the \textsc{mpmath} Python library, which our \textsc{mrpy} library utilises.}. Fortunately, $\Gamma$ admits a stable recurrence relation which can be used to efficiently calculate the function with $z<0$, in terms of only positive values, so long as $|z|$ is not too large (in our case, it should not exceed 3 very often). We present details of the mathematical derivation and numerical implementation of this relation for our work in Appendix \ref{mrp:app:recurrence}.

\subsection{Normalisation}
While in principle, $A$ can be set arbitrarily to fit given data, there are three values of interest. Firstly, to ensure the MRP is a valid statistical distribution defined on the support $(\mmin,\infty)$, one can set 
\begin{equation}
\label{mrp:eq:pdf_norm}
A = A_1 = 1/ \Hs \Gamma(z,x).
\end{equation}

Alternatively, to normalise the MRP so that its magnitude is equivalent to the expected number density of objects, $\frac{dn}{dm}$, we may make the (standard halo model) assumption that all CDM density is contained within halos (including subhalos) at some scale \citep{Cooray2002}. In this case, we force the integral of the mass-weighted function over all masses to converge to the known matter density, $\Omega_m\rho_\text{crit}$.
We find that 
\begin{equation}
    A = \Arc  = \Omega_m\rho_\text{crit}/k(\thp)
\end{equation}
where
\begin{equation}
    \label{mrp:eq:k}
    k(\thp) = \Hs^2 \Gamma(z_1),
\end{equation}
which is defined for $\alpha\geq-2$, as per previous discussion. 

We note that in practice, setting the normalisation in this way often leads to systematic errors with magnitude highly dependent on the value of $m_{\rm min}$, and thus is not suitable when fitting data. This problem is similar to that faced by the PS formalism \citep{Jenkins2001}.

A third value of interest arises from matching the expected total number of variates in the distribution to that observed. For $N$ haloes in a volume $V$, we have
\begin{equation}
	\label{mrp:eq:norm_to_counts}
	A = A_N = A_1 N/V.
\end{equation}

Table \ref{mrp:tab:properties} lists some further properties and derived quantities of MRP. Each of these is defined in our accompanying python code, \textsc{mrpy}.

\begin{deluxetable}{ll}
\tabletypesize{\footnotesize} 
\tablecolumns{2} 
\tablewidth{0pt} 
\tablecaption{Basic properties and derived quantities of MRP}
\tablehead{\colhead{Property} & \colhead{Value ($z\equiv z_0$)}} 
\startdata
 $n(>m|\thp)$ & $q(\thp,m)$ \\ 
  $n(>m_l, < m_u|\thp)$ & $ q(\thp,m_l)-q(\thp,m_u)$ \\ 
 $\rho(>m|\thp)$ & $A \Hs^2\Gamma(z_1,x)$ \\ 
 Mode & $\begin{cases} \mmin & \alpha < 0 \\ 	\Hs \left(\frac{\alpha}{\beta}\right)^{1/\beta} & \alpha > 0 \end{cases}$ \\
 $n^{th}$ Raw Moment & $\displaystyle \Hs^n \frac{\Gamma(z_n,x)}{\Gamma(z_0,x)}$ \\
 
 Mean & $\displaystyle \Hs \frac{\Gamma(z_1,x)}{\Gamma(z,x)}$ \\
 $n^{th}$ Central Moment & $\displaystyle \Hs^n \sum_{k=0}^n (-1)^{n-k} \frac{\binom{n}{k} \Gamma(z_1,x)^{n-k} \Gamma(z_k,x)}{ \Gamma(z_0,x)^{n+1-k}}$ \\
 
 Variance & $\Hs^2 \left(\frac{\Gamma(z+2/\beta,x)}{\Gamma(z,x)} - \left(\frac{\Gamma(z+1/\beta,x)}{\Gamma(z,x)}\right)^2\right)$ \\
 Log. Mass Mode, $\displaystyle \mathcal{H}_T$ & $\Hs\sqrt[\beta]{z+1/\beta}$
\enddata
\tablecomments{Properties and derived quantities of MRP. The mass mode is the logarithmic mass bin with the highest mass density.}
\label{mrp:tab:properties}
\end{deluxetable}

\subsection{Re-parameterisations}
\label{mrp:sec:repar-compare}
Re-parameterisations of the MRP are available which can reduce the covariance between the parameters \citep{LagosAlvarez2011,Lawless1982}. We explore some of these additional parameterisations in more detail in Appendix \ref{mrp:sec:repar}. We only comment here that while these may be beneficial in attaining parameter fits, we choose to use the parameterisation in Eq. \ref{mrp:eq:g} because the parameters $\thp_i$ have well defined characteristics in this form: $A$ is the normalisation, $\hs$ is the turnover mass, $\alpha$ is the low-mass power-law slope, and $\beta$ controls the sharpness of the exponential cut-off at high mass. 

We note here that it is possible to formulate the MRP in terms of the logarithmic mass mode $\mathcal{H}_T$ directly, where $\mathcal{H}_T$ is the mass at which
\begin{equation}
	\frac{d}{dm} mg(m|\thp) = 0,
\end{equation}
and that this parameter is closely related to $\Hs$ (the exact relation is given in Table \ref{mrp:tab:properties}). Conceptually, $\mathcal{H}_T$ is the `typical' mass of a halo for a given distribution, and we expect it to be less correlated with the other parameters than $\Hs$. It may be beneficial in future studies to adopt this parameterisation, but we do not here for the sake of simplicity and ease of comparison to the Schechter function.

We also note that the Schechter function affords the same relation, though since $\beta=1$ and $\alpha$ is generally close to $-1$ for luminosity functions, $\Hs$ is often very close to $\mathcal{H}_T$.

\section{Parameter Estimation with MRP}
\label{mrp:sec:fitting}


In this section we outline a general method for parameter estimation with the MRP. The key quantity in Bayesian estimation is the log-likelihood, which may be minimized using down-hill gradient optimization methods, or traced with Markov-Chain Monte Carlo (MCMC). 
We follow the usual method of defining the log-likelihood common in cluster cosmology studies \cite[eg.][]{Bocquet2016,DeHaan2016}.




A realised sample of haloes will be drawn from the MRP distribution within a certain volume. 
In this work, we incorporate into this `volume' the effects of a selection function -- that is, the volume is an effective volume at a given locus of the halo properties. 
The halo properties are not necessarily univariate -- i.e. we may be interested in more than just mass -- however in this work we will restrict ourselves to this univariate distribution.
The effective volume is also potentially dependent on the model parameters, $\thp$. 
Thus we specify the effective volume as $V(m|\thp)$, noting that the selection function for a simulation may be modelled as a simple step-function in $m$, which would incur the usage of $q(\thp,\mmin)$ as the number-density cumulant.

Let $\vec{x}$ be a set of observed properties of the haloes -- mass, velocity dispersion, luminosity, occupation number etc. -- and define $\rho(\vec{x}|m,\thp)$ as the probability that a halo of mass $m$ in the model with parameters $\thp$ is observed with the properties $\vec{x}$. 
Given an arbitrary binning of the data, and an occupation of those bins which is independent and Poisson-distributed, then going to the limit of zero-width bins, the likelihood of the parameters given a sample of $N$ haloes is
\begin{align}
	\label{mrp:eq:basic_lnl}
	\ln\mathcal{L}(\thp) = &-\int V(m|\thp) \phi(m|\thp)dm \nonumber \\
	&+ \sum_{i=1}^N \ln \int V(m|\thp)\phi(m|\thp)\rho_i(\vec{x}_i|m,\thp)dm.
\end{align}
Note that this formula does not include effects from cosmic variance or other systematic uncertainties.
We briefly discuss how one may incorporate such effects in Appendix \ref{mrp:app:cosmic_variance}, but we do not include them in any analyses in this work.
We expect that these effects will be dominant only for deep small-area surveys (i.e. those approaching a `pencil-beam' configuration), and we focus on larger-area surveys (or indeed large-volume simulations) in our examples.
Care should be taken in any application as to whether this assumption is valid.

%


We note a couple of special cases here as examples. Firstly, in the case of the MRP where we directly measure the precise mass of a halo (eg. from an $N$-body simulation), we have $\rho(\vec{x}|m\thp) = \delta(m'-m)$ (with $m'$ the observed mass), and we are left with
\begin{equation}
	\label{mrp:eq:exact_mass_lnl}
	\ln\mathcal{L}(\thp) = -\int Vg dm + \sum_{i=1}^N \ln\left( V(m'_i)g(m'_i)\right),
\end{equation}
where we have dropped the explicit dependence on $m$ and $\thp$ for each function for clarity, and shall do so hereafter unless it promotes ambiguity.

Secondly, if the selection function is truly a step function (a reasonable approximation in an $N$-body simulation) with a step at $\mmin$, up to volume $V_0$, then we have
\begin{equation}
	\label{mrp:eq:exact_mass_step_lnl}
	\ln\mathcal{L}(\thp) = -V_0 q(\thp,\mmin) + \sum_{i=1}^N \ln \left(V_0 g(m'_i)\right).
\end{equation} 

In practice, typically neither of these special cases are applicable, and the general Eq. \ref{mrp:eq:basic_lnl} must be used. 

\subsection{Acceleration Schemes}
It has been widely realised that the bottleneck in the calculation of Eq. \ref{mrp:eq:basic_lnl} lies in the evaluation of $N$ integrals in the second term \citep[eg.][]{DeHaan2016}.
In fact, the situation as presented here is already somewhat ideal -- we have assumed that each halo mass is uncorrelated with all other haloes. 
If this is not the case, then the $N$ 1-dimensional integrals become a single $N$-dimensional integral, which is clearly numerically intractable\footnote{In such cases, specifying each ``true" mass as a hyper-parameter and forming the model as a hierarchical Bayesian model is a useful way forward. Such models can be reasonably efficiently evaluated using advanced Monte-Carlo step methods such as Hamiltonian Monte Carlo, as implemented for example in the \textsc{stan} language. One requires that the likelihood be analytic to use \textsc{stan}, which provides another motivation for using a direct form such as the MRP.}, and we shall not pursue such difficulties here.

One interesting way to mitigate this bottleneck is to follow the \textit{fit-and-debias} method of \citet{Obreschkow2017}.
This method proposes a new likelihood, based on the most likely posterior mass for each observation:
\begin{equation}
\label{mrp:eq:fit_and_debias}
\ln\mathcal{L}(\thp) = -\int V \phi dm + \int \ln \left(V\phi\right) \sum_{i=1}^N \bar{\rho}_i(\vec{x}_i) dm,
\end{equation}
where 
\begin{equation}
	\bar{\rho_i} = \frac{V\phi\rho_i(\vec{x}_i)}{\int V\phi\rho_i(\vec{x}_i) dm}.
\end{equation}
This likelihood has a maximum at the input parameters $hat{\theta}$ iff $\bar{\rho}$ is defined to be constant w.r.t $\thp$, and is evaluated at $\hat{\theta}$. 
Thus, this method follows an iterative approach in which first $\rho_i$ is evaluated at a given estimate $\theta_0$, then the likelihood is maximized to yield a better estimate, $\theta_1$, and so on until convergence. 

In this approach, the full $N$ integrals need only be calculated once every method iteration, rather than on every likelihood evaluation. 
This significantly speeds up calculations for which these integrals are the bottleneck.

One must be aware, however, that the likelihood used here, while having the same maximum point as Eq. \ref{mrp:eq:basic_lnl}, is not the same likelihood. 
Thus, the parameter errors derived from the likelihood are not necessarily accurate, and one should resort to the proper likelihood to evaluate these.

\subsection{Expected Parameter Covariance}
\label{mrp:sec:expected_cov}
The errors on parameter estimates, given a sample of data, are most robustly evaluated using the chains from an MCMC sampling of the posterior.
However, a more efficient estimate of the covariance of parameters, assuming a Gaussian posterior at the maximum likelihood estimate (MLE), can be gained from the Hessian, as $-(H(\thp))^{-1}$.
Furthermore, in the process of model-building or survey-construction, one may be more interested in the \textit{expected} covariance on their parameters, rather than the covariance actually obtained from a given set of data.

The expected Hessian gained from sampling a model $\phi$ with volume $V$ and uncertainties $\rho$ is (cf. Appendix \ref{mrp:app:expected})
\begin{equation}
	\label{mrp:eq:expected_hessian}
	H(\thp) = - \int \frac{\left.\frac{\partial n(\vec{x})}{d\thp_i}\right|_{\hat{\theta}}\left.\frac{\partial n(\vec{x})}{d\thp_j}\right|_{\hat{\theta}}}{n(\vec{x}|\hat{\theta})} d\vec{x},
\end{equation}
where $n(\vec{x}) = \int V\phi\rho(\vec{x})dm$.

This expression for the Hessian is beneficial, as it involves only first derivatives of $n(\vec{x})$, which improves numerical efficiency and accuracy over second derivatives.
Interestingly, Eq. \ref{mrp:eq:expected_hessian} is analytically solvable in the simplest case of the MRP with precisely observed masses.
We provide this solution in Appendix \ref{mrp:app:special_case_3}.

\section{Accuracy of the MRP}
\label{mrp:sec:accuracy}
The Schechter function is (usually) a reasonable fit to luminosity or stellar-mass data, up to the uncertainty in the data itself. In the case of the MRP, we have a wealth of information, in the form of first-principles simulations and fits to such simulations using the PS formalism. Thus, our first task is to measure the accuracy of the MRP with respect to the precision of our current knowledge of the theoretical HMF. Clearly, the MRP will be an imperfect model; it will lose information in the reduction of a simulation to 5 parameters. However, we wish to test whether the MRP is an adequate approximation to theoretical HMFs up to the uncertainty both inherent in the theory, and in relevant datasets.

We approach this in two steps. First, we perform simple comparisons to HMF fits using the PS formalism, allowing us to quickly explore parameter space and assess the accuracy of MRP in different regimes. Secondly, we compare the MRP to the output of a full $N$-body simulation, comparing the residuals to those from the best fit using the PS formalism. Our aim is not to show that the MRP performs better than the PS formalism, but that it performs adequately for use with real data.

Throughout this paper we adopt the fiducial cosmology of \citet{PlanckCollaboration2014} (hereafter P13)\footnote{Results for the more recent 2015 data release are almost identical, and the choice of P13 is driven by its use in simulations in \S\ref{mrp:sec:fitnu2}}. Likewise, we adopt the fit of \citet[hereafter T08]{Tinker2008} as our fiducial mass function.

\subsection{Comparison to PS-type fits}
\label{mrp:sec:compareEPS}
\begin{figure*}
\centering
\includegraphics[width=\linewidth]{\mypref 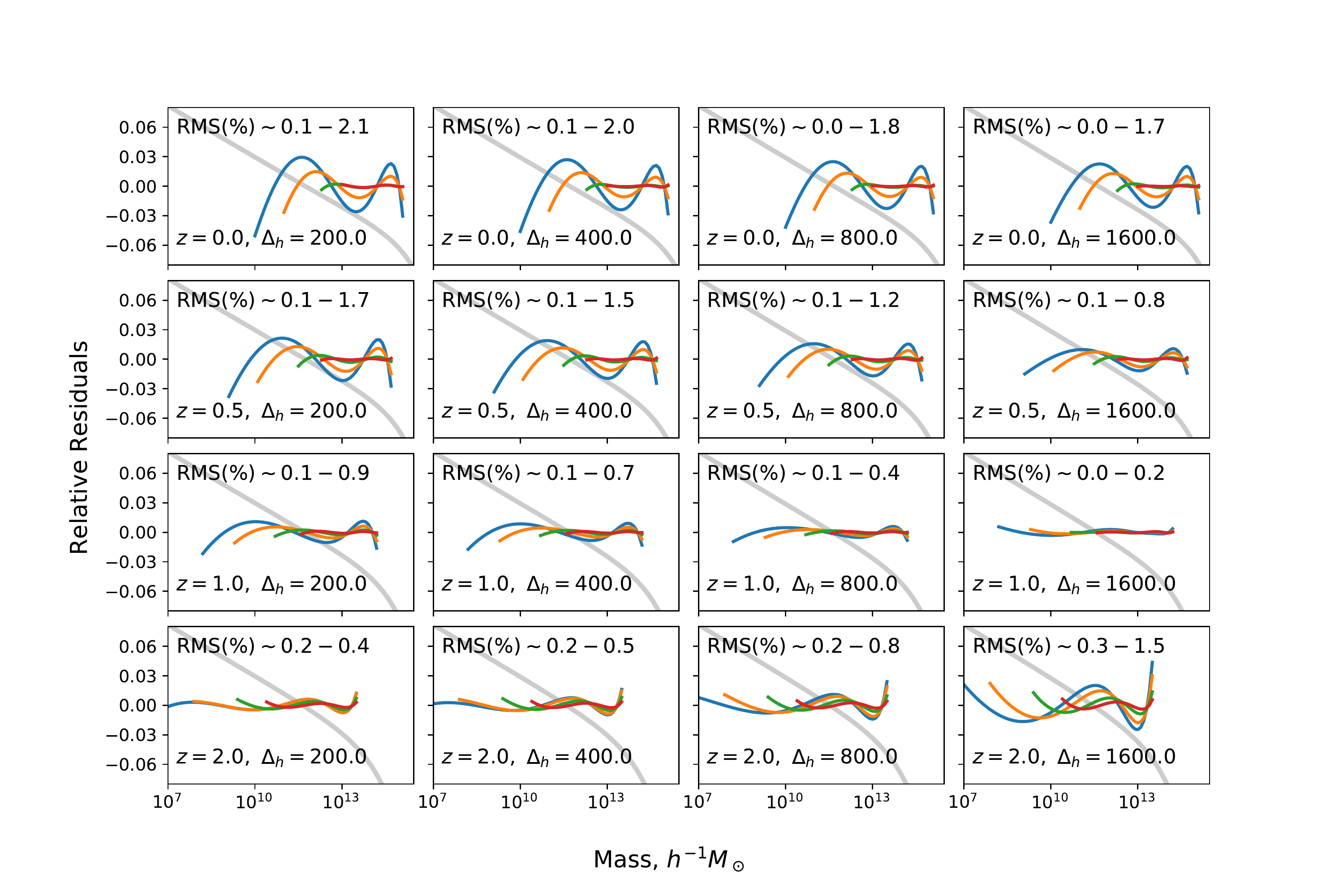}
\caption[Comparison of MRP to Tinker fitting function]{Comparison of MRP form to T08 over several redshifts and over-densities. Each column represents an over-density corresponding to those reported in T08. Each row represents a redshift, up to $z=2$. The four lines in each represent a fit over a different mass range. The grey curves in the background of each panel show the log-scale HMF in arbitrary units, in order to show the kind of shape that the MRP was fit to in each case. The rms deviation as a percentage is shown for the best and worse case in each panel. The rms is better than 2\% over all over-densities for $z\leq 1$, and is a maximum of 2\% for $z=0$ and $\Delta_h=200$. Expectedly, residuals are worse for fits over wider mass ranges.}
\label{mrp:fig:againstTinker}
\end{figure*}

The fit of T08 is very commonly employed in observational studies \cite[eg.][]{PlanckCollaboration2016,Pacaud2016}, since it provides an accurate fit over redshift (up to $z=2$) and also for varying spherical over-densities. Using a simple $\chi^2$-minimization in log-log space, we fit T08 for a range of redshifts and halo over-densities, as well as varying lower mass limits.
The mass range for each is calculated between constant limits on the mass variance, so as to probe the interesting part of the HMF for each combination of parameters.
The limits are $\sigma_{\rm max} = [4, 3, 2, 1.5]$ and $\sigma_{\rm min} = 0.5$.
The T08 HMFs were produced with the \textsc{hmf} Python code, v3.0.2 \citep{Murray2013c}. Results are plotted in figure \ref{mrp:fig:againstTinker}.

The goodness-of-fit of the MRP varies across redshift and over-density, but overall even in the worst-case scenario the rms deviation over the fitted mass range is about 2\% (for $z=0$ and $\Delta_h=200$, where $\Delta_h$ is the over-density with respect to the mean matter density $\rho_m$). The approximate uncertainty in T08 is reported as 5\%, so in a global sense, we find the MRP to be a good approximation.

The variation of the residuals with redshift is such that they generally tighten towards higher redshift.
This can be ascribed to the fact that the mass function turnover becomes less marked at these higher redshifts -- the function is close to a steep power-law. 
This is seen by the background grey curves which represent log-scale HMFs of arbitrary units, to which each MRP was fit.
Since the transition from power-law to exponential cut-off is the most difficult regime for the MRP to model, it excels when this transition is muted.
 An increased over-density criterion adversely affects the residuals at high redshift, but has the opposite effect at low redshift. This can be attributed to the fact that a change in over-density is predominantly a non-linear shift of the HMF on the mass-axis. The primary resulting change in the HMF is the power-law slope at low redshift, but the position of the cut-off at high redshift. Thus the explanation for the increased residuals is similar to that of its variation with redshift. In effect, the MRP performs most poorly around the transition from power-law to exponential cut-off, but is still typically within 5\% over a large parameter space. In addition, it is difficult to ascertain whether the residuals above the transition are robust, since the T08 fit itself is more poorly constrained on these scales due to the inherent Poisson noise and cosmic variance in simulations.

\begin{figure*}
	\centering
	\includegraphics[width=\linewidth]{\mypref 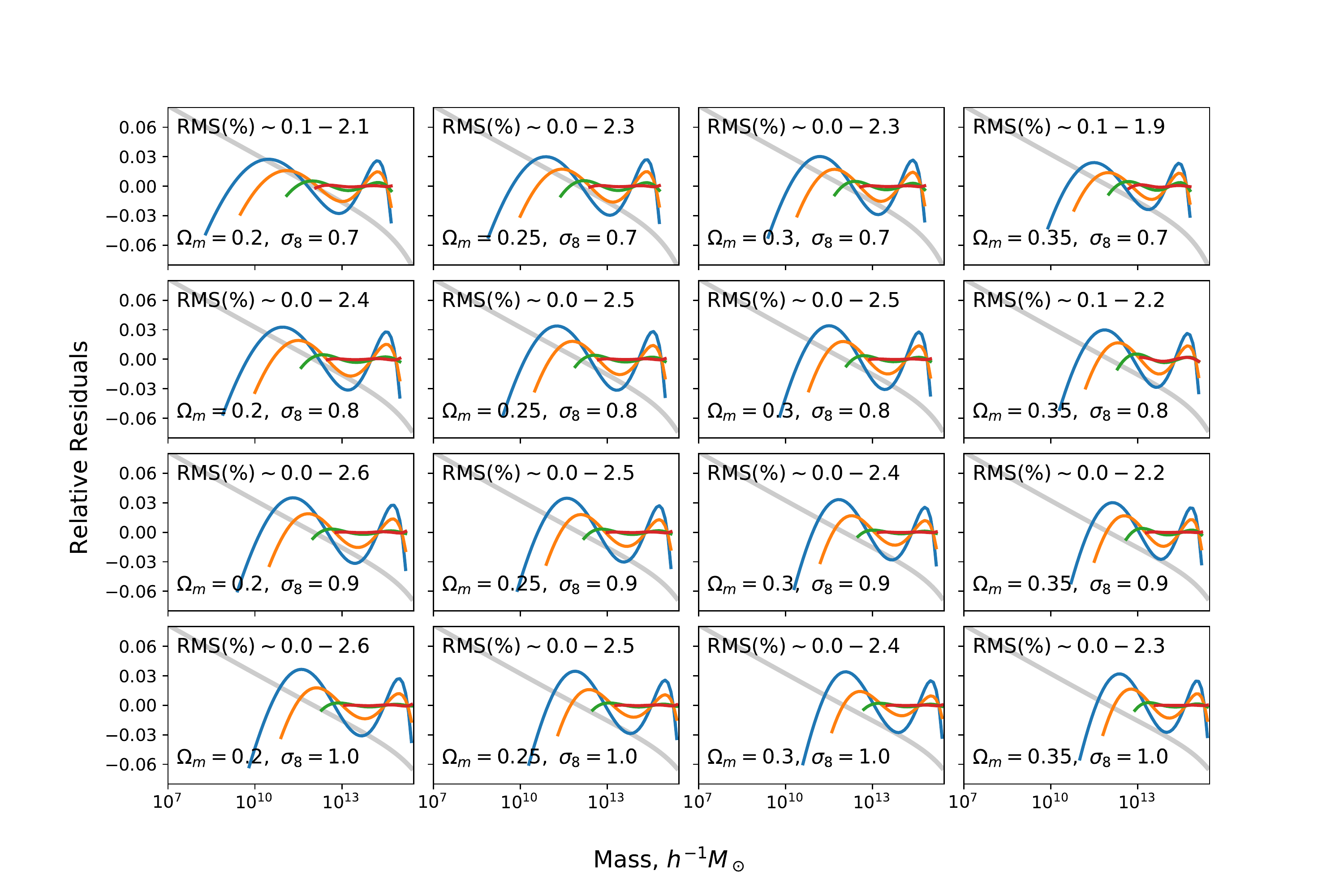}
	\caption[Comparison of MRP to Warren fitting function]{Comparison of MRP form to the fit of \citet{Warren2006} over several values of $\Omega_m$ and $\sigma_8$. Everything is the same as Fig. \ref{mrp:fig:againstTinker}, except that here each column represents a value of $\Omega_m$, and each row a value of $\sigma_8$. The rms is better than $\sim 2.6$\% over all parameters, and is a maximum of 2.6\% for $\Omega_m=0.2$ and $\sigma_8=1.0$}
	\label{mrp:fig:againstWarren}
\end{figure*}

In order to assess whether the MRP form is flexible enough to accommodate various cosmologies and halo finders, we provide Fig. \ref{mrp:fig:againstWarren}, which is very similar to Fig. \ref{mrp:fig:againstTinker}, except that here the fitting function is that of \citet{Warren2006} -- a form fit to Friends-of-Friends (FOF) halos -- and the columns and rows represent $\Omega_m$ and $\sigma_8$ respectively.
We firstly find that large changes in cosmology have little effect on the ability of the MRP to describe the HMF -- the difference in rms between all curves over the same range is less than 1\%. 
Secondly, we find that the fit to an FOF-based HMF is slightly worse than that to the SO-based T08 function, but only by $\sim 0.5$\% (we compare to the top-left panel of Fig. \ref{mrp:fig:againstTinker}, which should be the most similar to the curves in this figure).
Thus we conclude that the MRP is flexible enough to fit HMFs from various cosmologies, redshifts, halo-finders and halo-definitions with roughly equivalent accuracy -- and that to within the quoted uncertainties from PS-type fits.

The most concerning feature of the residuals is an oscillatory pattern in the power-law region of the distribution. This oscillatory pattern is expanded as the lower mass limit is decreased, indicating that it is a model deficiency. We do note however that similar oscillatory residuals are seen in standard PS formalism fits to $N$-body simulations.
It is of interest to what extent this feature is a loss of information in MRP as compared to the true physical distribution. This question is best answered by observing residuals of fits to simulations, to which we now turn.

\subsection{Comparison to $N$-body Simulations}
\label{mrp:sec:fitnu2}
The simplest realistic halo catalogues arise from DM-only simulations. Such catalogues contain high-resolution, fully non-linear determinations of the HMF, completely independent of the PS framework assumptions. This enables comparisons of the fidelity of the MRP versus the PS formalism without any possible biases from the PS modelling itself. 


In this study, we use the publicly available halos from the New Numerical Galaxy Catalog ($\nu^2$GC, \citealt{Ishiyama2015}; hereafter I15). These simulations were run with the P13 cosmology, in several box-sizes and mass resolutions. Specifically, we use $\nu^2$GC-M and $\nu^2$GC-H1, which when taken together probe a volume of 560 $h^{-3}$Mpc$^3$ down to halos of mass $2.75\times10^9 h^{-1}M_\odot$. As such, it spans the halo mass range relevant for group catalogue studies. Details of the simulations are found in Table \ref{mrp:tab:n2gc}. In particular, we note that though the original catalogues contain haloes defined down to 40 particles, we fit only to those with 100 particles or more, to alleviate the effect of Poisson noise on the masses of these highly influential haloes. Furthermore, the box size of $\nu^2$GC-M is small enough to make the highest-mass objects quite rare, and thus dominated by cosmic variance and Poisson noise. We set upper limits of 7$\times10^{14}h^{-1}M_\odot$ and 2$\times 10^{13}h^{-1}M_\odot$ on the mass ranges respectively.

In order to test the robustness of the MRP to different halo definitions, we use both Friends-of-Friends (FOF) and Spherical-Overdensity (SO) haloes identified in each simulation.
The SO haloes are defined via the critical overdensity, $\Delta_h = 200  \rho_c$ using the \textsc{rockstar}\footnote{\url{https://bitbucket.org/gfcstanford/rockstar}} code \citep{Behroozi2011}. Note that \textsc{rockstar} is a 6D FOF halo-finder, but does produce SO halos as part of its output.

\begin{deluxetable*}{lllllll}
\tablecolumns{7}
\tablewidth{0pt}
\tablecaption{Details of halo abundances in the $\nu^2$GC boxes used in this study.}
\tablehead{Name & $N$ & $L,\ [h^{-1}{\rm Mpc}]$ & $m_p,\ [h^{-1}M_\odot]$ & \# Halos & $M_\text{min},\ [h^{-1}M_\odot]$ & $M_\text{max},\ [h^{-1}M_\odot]$ }
\startdata
$\nu^2$GC-M& 4096$^3$ & 560 & 2.2$\times10^8$ & 53$\times10^6$ & 2.2$\times10^{10}$ & 7$\times10^{14}$ \\


$\nu^2$GC-H1& 2048$^3$ & 140 & 2.75$\times10^7$ & 5.5$\times10^6$ & 2.75$\times10^9$ & 2$\times10^{13}$

\enddata
\tablecomments{$M_{\rm min}$, as given by the public catalogues, is set by a minimum of 40 particles per halo. For our fits, we prefer to use a minimum of 100 particles per halo. Note that this changes the number of halos that are useable for our fits as compared to the number shown here. The SO halos have exactly the same parameters except number of halos.}
\label{mrp:tab:n2gc}
\end{deluxetable*}

We estimate the parameters of the MRP via the likelihood in Eq. \ref{mrp:eq:exact_mass_step_lnl}. 
However, some care must be taken when simultaneously analysing multiple datasets with different volumes. In this case, we calculate the log-likelihood for each data set with the same shape parameters, but altering the normalisation so that it is relatively correct for each dataset. More specifically, given $n$ datasets, each with $N_j$ haloes in a volume $V_j$, we use
\begin{equation}
	\ln \mathcal{L} = \sum_{j=1}^n -V_j q(\thp,m_{\rm min, j}) + N_j \ln V_j + \sum_{i=1}^{N_j} \ln g(\thp,m_i).
\end{equation} 

In our case, this means that over most scales, the $\nu^2$GC-M box is highly weighted, but at the smallest scales which it does not probe, the H1 box is effective.  
In general, since we have manually truncated the distribution at the high-mass end, we should amend $q$ so that it is correctly calculated between the mass limits. However, due to the
 rapid decline of the distribution, such an amendment has a very small effect, and in this case we neglect it.

The parameter estimation can be performed either with simple likelihood maximisation (via downhill-gradient methods), or MCMC. For illustration purposes we use MCMC in this case. 
The MCMC sampling used the \emcee\footnote{\url{http://dan.iel.fm/emcee/current/}} python package \citep{Foreman-Mackey2012}, in which we used 50 ``walkers" each for 200 steps of burn-in, and 500 retained steps, for a total of 25,000 samples. We begin the chains in a small ball around the optimization solution, as recommended in \cite{Foreman-Mackey2012}. We verify convergence with the Gelman-Rubin test \citep{Gelman1992}, ensuring that the \textit{potential scale reduction factor} $\hat{R} < 1.1$ for all parameters.

\begin{figure*}
	\centering
	\includegraphics[width=\linewidth]{\mypref 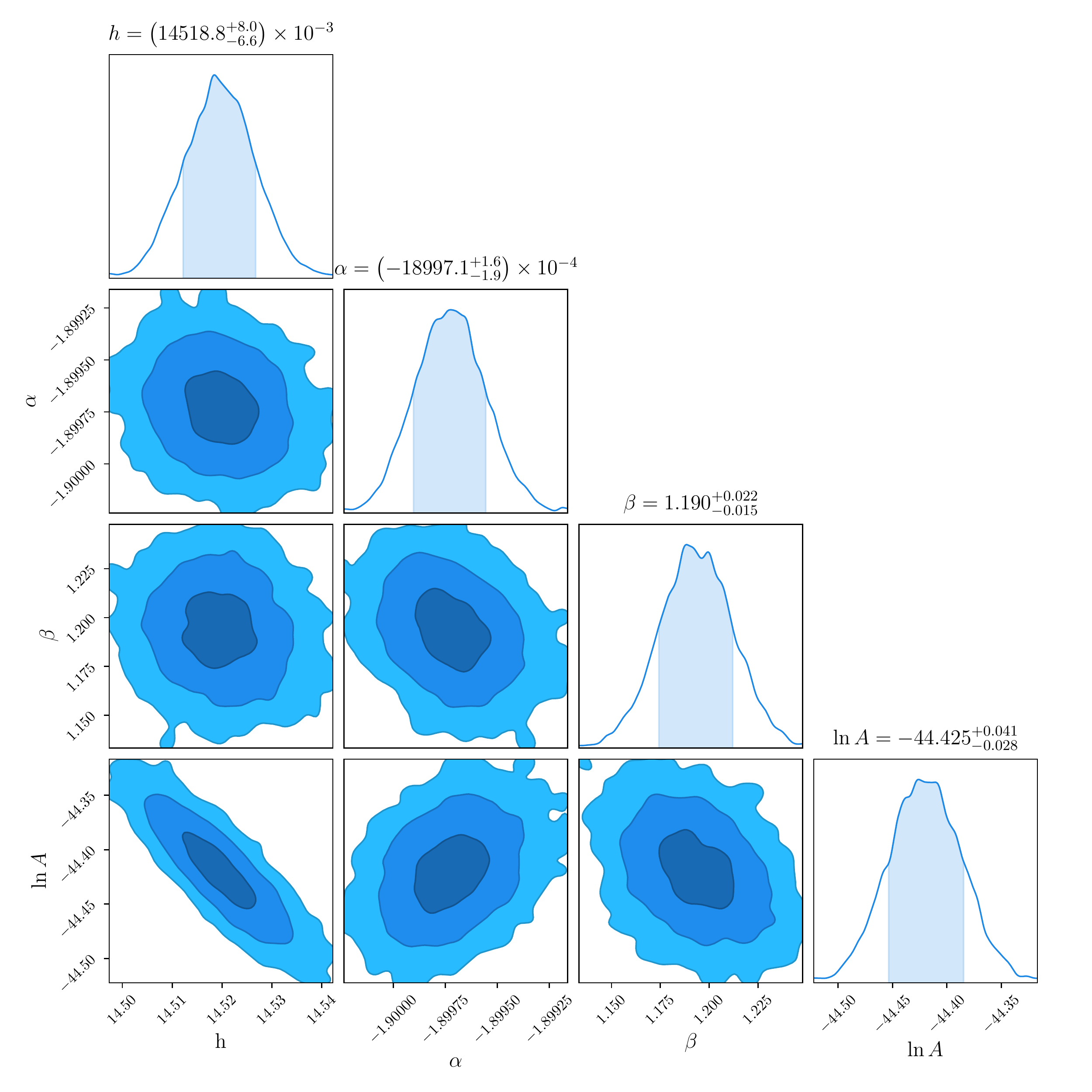}
	\caption[Joint posteriors of MRP parameters estimated simultaneously on two $\nu^2$GC simulations]{Posterior joint-likelihoods of the MRP parameters estimated simultaneously on both $\nu^2$GC simulations, using FOF halos. All contours display high degrees of normality. The distributions are rather covariant, which can pose problems for standard sampling techniques.}
	\label{mrp:fig:joint-n2gc}
\end{figure*}

We plot the posterior joint-likelihoods of the parameters, for the FOF halos, in Figure \ref{mrp:fig:joint-n2gc}.
We stress that this is a high-resolution run, which will be poorly represented by realistic observed datasets. With that in mind, the joint posterior is a very good approximation to a multivariate normal distribution, which has the benefit of allowing for quick analyses by downhill-gradient methods, combined with analytic determination of the distribution around the solution via the Hessian (see Appendix \ref{mrp:app:jac_hess_po} for the exact equation for this quantity). There is a high degree of covariance, as expected, between $\ln A$ and $\hs$. 

Because of the high volume of data, the parameters are extremely well constrained, but their precision is not necessarily the best metric of performance. Since the underlying data does not perfectly trace the MRP form, the parameters are likely to be biased, and it is the residuals that are the most telling.  
The turnover-mass is around $10^{14.52} h^{-1}M_\odot$, and the mass mode is $\sim 10^{13.62} h^{-1}M_\odot$. We find a power-law slope of -1.9. We also find a cut-off parameter $\beta$ of 1.19, which is significantly sharper than a Schechter function (in which $\beta\equiv1$). 
The log-normalisation is -44.4. Of these, we expect the value of $\beta$ to be the most inaccurate, as it is principally determined by the high-mass haloes, of which there are very few.
 


\begin{figure*}
	\centering
	\includegraphics[width=\linewidth]{\mypref 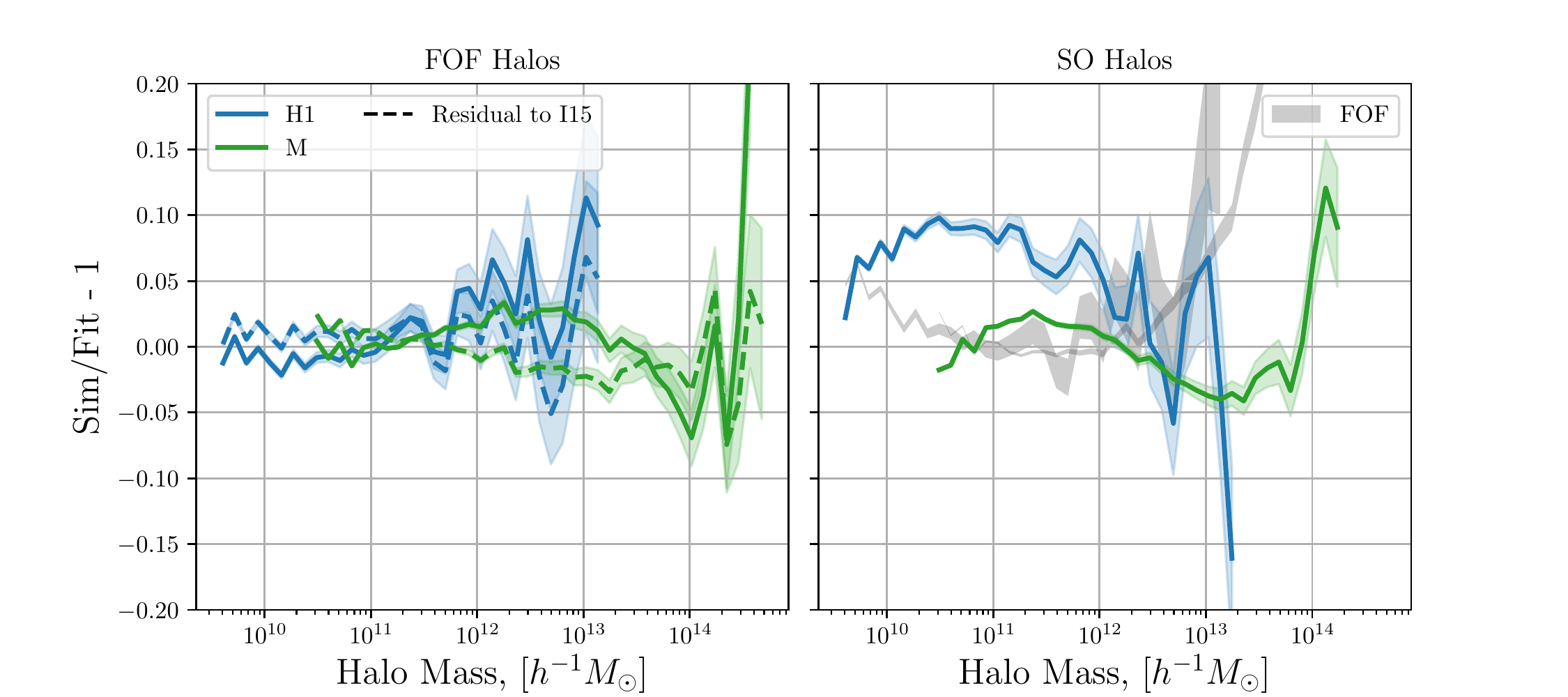}
	\caption[Residuals of MRP fit to $\nu^2$GC simulations]{Residuals of fits to $\nu^2$GC simulations. The solid lines show residuals from the suites of halos against the global MRP fit. The dashed lines show the I15 fit residuals to the same haloes. The left panel shows the FOF halo fit, while the right panel is normalised to the SO halo fit. Shaded regions indicate the 1-$\sigma$ uncertainty regions generated from the Poisson noise estimate. The grey shaded regions in the right-hand panel are the naive FOF binned HMFs normalised to the SO fit, to illustrate the magnitude of the difference between halo-finders.}
	\label{mrp:fig:n2gc-resid}
\end{figure*}

Figure \ref{mrp:fig:n2gc-resid} shows the residuals of this fit as thick solid lines (different colours representing different simulation boxes). 
The left panel shows the fit to the FOF haloes, while the right panel shows the fit to the SO halos.
In each case, the figure shows the naively binned HMF from the simulation as a ratio to the best fit. 

In addition, the right panel shows the binned FOF HMF as a ratio to the SO fit in grey, to illustrate the magnitude of the difference between the halo-finding schemes. 
The MRP fit is accurate to within 5 percent over the mass range $10^{9.7} - 10^{14} h^{-1} M_\odot$, which is a far smaller margin than the discrepancy between halo finders. 
This indicates that an arbitrary choice of halo definition is a greater loss of information than the approximation introduced by the MRP.

Also plotted in figure \ref{mrp:fig:n2gc-resid} are the residuals of the best-fit PS-type mass function from I15 (dashed lines), which was calibrated with the FOF halos (and therefore is only shown in the left panel).
The residuals are comparable in magnitude over much of the mass range, with the PS-type function providing a tighter fit  at $\sim 10^{12}-10^{13} h^{-1}M_\odot$.
At lower masses, the MRP is much more accurate than the PS-type fit. The reason for this difference in accuracy at different scales arises from at least two considerations. First, the MRP (and the PS HMF) are not perfect representations of the HMF, and therefore cannot capture the behaviour on all scales. Secondly, while it is likely that the I15 form was fit to binned counts in log-log-space, and therefore treats all scales equally, our likelihood implicitly fits the HMF in real-space, so that the low-mass haloes dominate the fit. Whether one fit is ``better" than the other is therefore based on the application at hand. Typically, for plots such as Figure \ref{mrp:fig:n2gc-resid}, fits in log-log-space will appear more accurate over a larger range.

The fit to the SO haloes performs relatively poorly, and this is due to the inconsistency between the two catalogues involved, as can be clearly seen in figure \ref{mrp:fig:n2gc-resid}.
It is unclear whether this is due to sample variance, or a miscalibration in the SO algorithm, but the results from these catalogues should be treated with caution.

The oscillation that was apparent in the fits to the PS formalism HMFs is still present, indicating that the MRP is not flexible enough to capture the details of the HMF distribution. However, clearly at some level this is also true of the PS formalism, which systematically over-predicts the low-mass end and under-predicts the high-mass end of the HMF in this simulation. The MRP is expected to perform more poorly than the PS formalism when the mass range is increased. This must be remembered when using the MRP for any fit -- the truncation scale(s) are an inherent part of the formula, and will to some extent affect the estimation of the other parameters.

Nevertheless, we see that MRP is able to empirically model the HMF over a sufficiently large mass range to be relevant for group catalogues. 



\section{Dependence on Physical Parameters}
\label{mrp:sec:dependence}
While it is envisaged that the MRP will generally be used as a standalone description of halo mass data, it is instructive to determine the dependence of the parameters $\thp \equiv (\Hs,\alpha,\beta,\ln A)$ on the more important physical parameters.

The parameter with the most influence over the HMF is the redshift, which through the growth rate affects the normalisation of the power spectrum, and is also explicitly modelled into the HMF fit in several recent studies (eg. our fiducial model of T08). Along with this, \citet{Murray2013} showed that the parameters $\Omega_m$ and $\sigma_8$ have the most significant effects across a broad range of masses. Thus, in this section we set out to model the dependence of the $\thp_i$ on $\vec{\phi}  = (\Omega_m, z,\sigma_8)$. 

We note that the choice of base model to fit to is somewhat arbitrary. Fits in the literature vary at the level of tens of percent across interesting mass ranges. Furthermore, there is a dichotomy between fits based on haloes found using FOF and SO methods, as we have seen. Given the inherently approximate nature of this dependency modelling, we opt to fit to a chosen fiducial model, and use the uncertainty of this fit (both intrinsically and with respect to other fits) to guide our judgement of the appropriateness of the model. 

Since we are interested in modelling both low and high redshifts, we take as our fiducial fit the form of \citet{Behroozi2013} (hereafter B13), which is an empirical modification of T08 that increases accuracy at high redshift (up to the Epoch of Reionization).

When calculating the HMFs for modelling the MRP parameters, we ensure that in each case the wavenumber range is wide enough to encapsulate all relevant information (see \S2 of \citet{Murray2013c} for a brief discussion of relevant limits on the product of the radii and wavenumber), and also that the resolution is high enough to avoid small oscillatory artefacts. 

\begin{figure}
    \centering
    \includegraphics[width=\linewidth]{\mypref 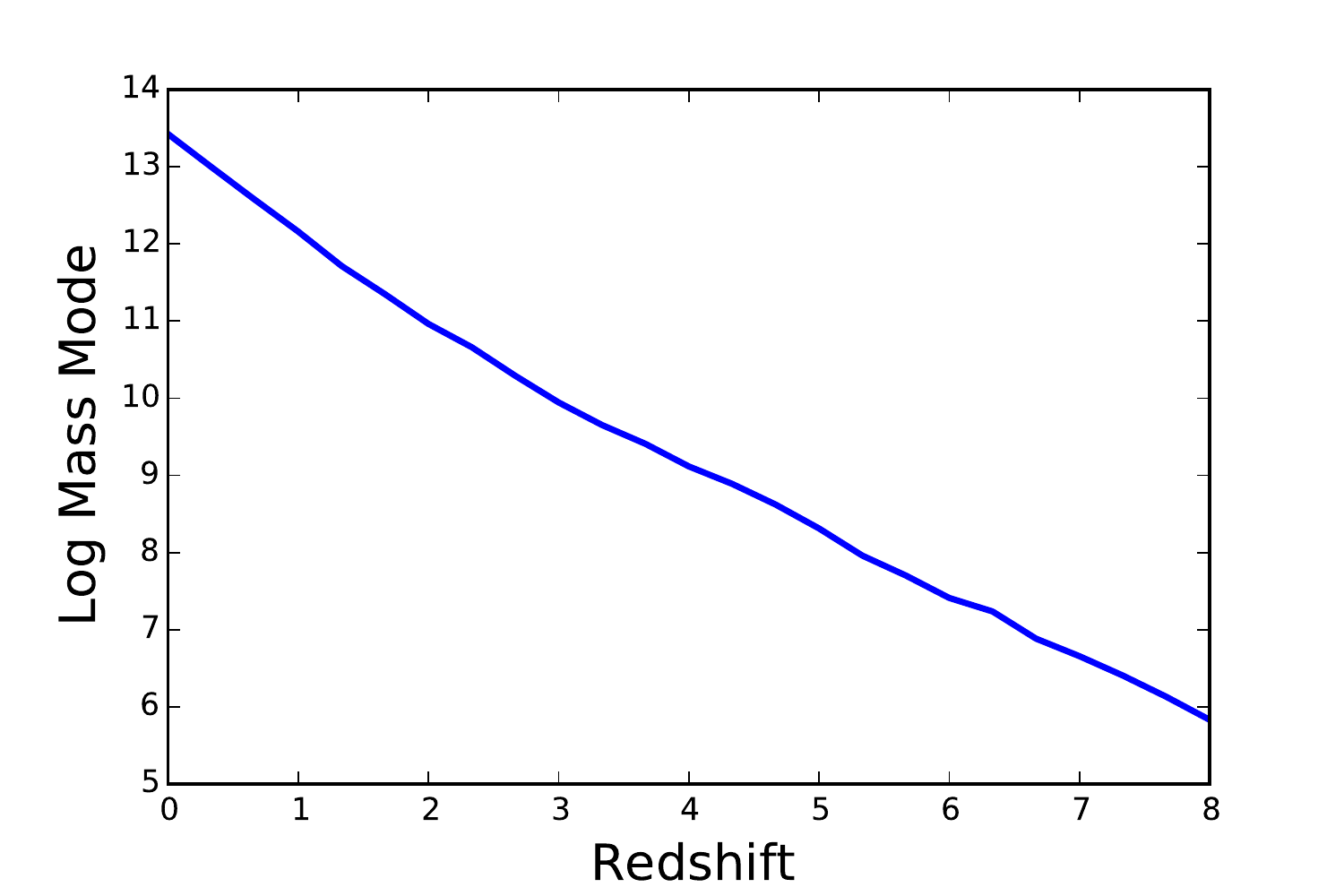}
\caption[Logarithmic mass mode $\mathcal{H}_T$ as a function of redshift.]{Logarithmic mass mode $\mathcal{H}_T$ as a function of redshift for the B13 HMF.}
\label{mrp:fig:log_mass_mode}
\end{figure}

An important consideration is the mass range over which to fit the MRP, as this choice affects the derived parameters (cf. Fig. \ref{mrp:fig:againstTinker}). In order to probe the domains of influence of all three shape parameters, and to provide a standardized system, we opt to define the mass range as constant with respect to the logarithmic mass mode $\mathcal{H}_T$. This value is calculated explicitly as the zero of the derivative of a quartic spline interpolation of the B13 HMF for each value in the sample of $\vec{\phi}$. Figure \ref{mrp:fig:log_mass_mode} shows $\mathcal{H}_T$ as a function of redshift for fiducial values of $\Omega_m$ and $\sigma_8$ (the dependence on these latter parameters is comparatively negligible). We have verified that $\mathcal{H}_T$ is within the valid mass range of B13 for all samples.

It must be urged that the fitted parameters can change significantly given a different truncation mass with all other parameters fixed. Specifically, if a smaller mass range is of interest, a more accurate fit can generally be attained (cf. Fig. \ref{mrp:fig:againstTinker}). Furthermore, an optimization of the MRP parameters for any input model will achieve a greater accuracy than using the models presented in this section (and doing so is not computationally expensive). However, the most important benefit of this section is the insight gained from determining the approximate relationship between physical and MRP parameters, and that remains relatively consistent with variable $m_\text{min}$. 

We produce 2000 samples, $\vec{\phi}_i$. The redshifts are drawn uniformly in log-space, so as to lightly increase the weight of lower redshifts. The cosmological parameters are drawn from the normal distributions presented in P13, so as to highly weight values which are more likely to be physically appropriate. In summary,
\begin{align*}
    \log (1+z) &\sim U(0,\log 9) \\
    \sigma_8 &\sim \mathcal{N}(0.829,0.012) \\
    \Omega_m &\sim \mathcal{N}(0.315,0.017).
\end{align*}

A value $\thp_i$ is fit to each model using $\chi^2$-minimisation in log-log space (identical to the procedure for producing Fig. \ref{mrp:fig:againstTinker}).
To find an optimal parameterisation for $\thp$ as a function of $\vec{\phi}$, we use the symbolic regression program \textsc{eureqa}\footnote{\url{http://www.nutonian.com/products/eureqa-desktop/}} \citep{Schmidt2009,Schmidt2014}. This program iterates over models using a genetic algorithm, with constraints set by the user, to find underlying trends in input data. To assist the convergence of the program, we first fit a one-dimensional model in redshift only, which is the most effective parameter. Using this model as a baseline, we introduce $\Omega_m$ and $\sigma_8$ for a full 3D parameterisation. Each run of \textsc{eureqa} produces a series of potential parameterisations of varying complexity and we choose models that offer a balance between accuracy and simplicity.

\begin{figure}
\centering
\includegraphics[width=\linewidth, trim=1.5cm 0.5cm 1.5cm 0.5cm,clip]{\mypref 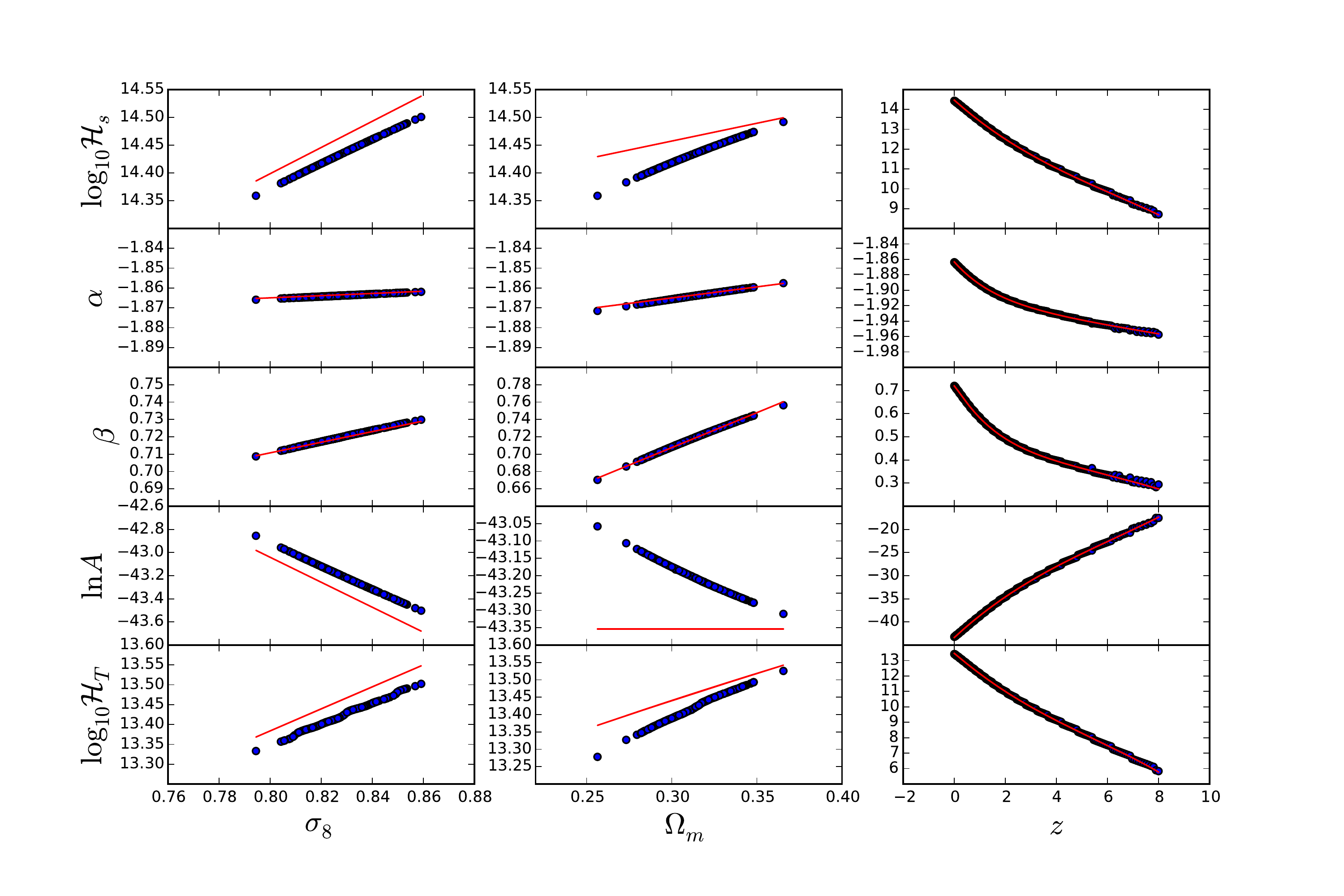}
\caption[Independent physical dependence of MRP parameters]{Dependence of $\thp_i$ (rows) on $\vec{\phi}$ (columns). The bottom row shows the logarithmic mass mode, $\mathcal{H}_T$. Blue markers show best-fit values derived from $\chi^2$-minimization on the B13 HMF. Red lines show the \textsc{eureqa}-derived models from table \ref{mrp:tab:3d}. The lower effectiveness of the cosmological parameters renders their independent models visually poorer, especially in the case of the normalisation.}
\label{mrp:fig:param_dep}
\end{figure}

We present our chosen models in table \ref{mrp:tab:3d}. Each model is normalised and scaled to its mean and standard deviation, highlighting the primary dependencies amongst the parameters. A more intuitive presentation of the models can be found in figure \ref{mrp:fig:param_dep}, in which the univariate dependencies of $\theta_i$ on $\phi_j$ are shown. Here, $\sigma_8$ and $\Omega_m$ are varied at $z=0$, and otherwise take their fiducial values.

\begin{deluxetable*}{lccc}
	\tabletypesize{\footnotesize} 
	\tablecolumns{4} 
	\tablewidth{0pt} 
	\tablecaption{Table showing formulae for the physical dependence of the MRP parameters}
	\tablehead{\colhead{Parameter} &  Formula, $\mu + \sigma f$ & $\mu$ & $\sigma$} 
	\startdata
	$\hs$ & $\displaystyle \begin{aligned} 0.058562 &+ 1.4394\sigma_8 + 0.39111\Omega_m + 0.11159\sigma_8 z \\ &+ 0.056010z^2+0.42444\sigma_8\Omega_m z - 0.90369z - 0.0029417z^3\end{aligned}$ & 12.214 & 1.6385 \\
	$\alpha$ & $\displaystyle 2.6172\Omega_m + 2.06023\sigma_8 + 1.4791\times 2.2142^{\Omega_m} 0.53400^z - 2.7098 - 0.19690z$ & -1.9097 & 0.026906 \\
    $\beta$ & $\displaystyle 7.5217\sigma_8\Omega_m - 0.18866 - 0.36891z - 0.071716\times 0.0029092^z - 3.4453\Omega_m z 0.71052^z$  & 0.49961 & 0.12913  \\
    $\ln A$ & $\displaystyle z + 0.0029187z^3 - 0.11541 - 1.4657\sigma_8 - 0.055025z^2 - 0.24068\sigma_8 z - 0.33620\Omega_m z$ &-33.268 & 7.3593
	\enddata
	\tablecomments{Parameterisations of $\thp_i$ as functions of $\mathbf{\phi}$. All parameterisations are expressed as rescaled by the mean and standard deviation of the input data, which highlights the typical values and associated sensitivity of each MRP parameter.}
	\label{mrp:tab:3d}
\end{deluxetable*}

In particular we note that across the board, the most effective parameter is $z$, with which all three MRP parameters are anti-correlated. The MRP parameter most sensitive to the physical parameters is $\beta$, followed by $\Hs$, with $\alpha$ changing very slowly (mainly with redshift). 

The visually poor performance of the models over independent cosmological parameters is due to the low effectiveness of these parameters. In particular, the normalisation has no dependence on $\Omega_m$ at $z=0$, which is clearly not the case in detail. This suggests that the ``automatic" fits from \textsc{eureqa} can be improved.

Summarily, the redshift dependence of the parameters indicates a crossover between two behaviours. At low redshifts there are two regimes -- a low-mass power-law tail, and a high-mass cut-off region. At higher redshifts, the high-mass cut-off moves to become so far beyond $\mathcal{H}_T$ that the function resembles a power-law over all relevant scales.

\begin{figure}
\centering
\includegraphics[width=\linewidth]{\mypref 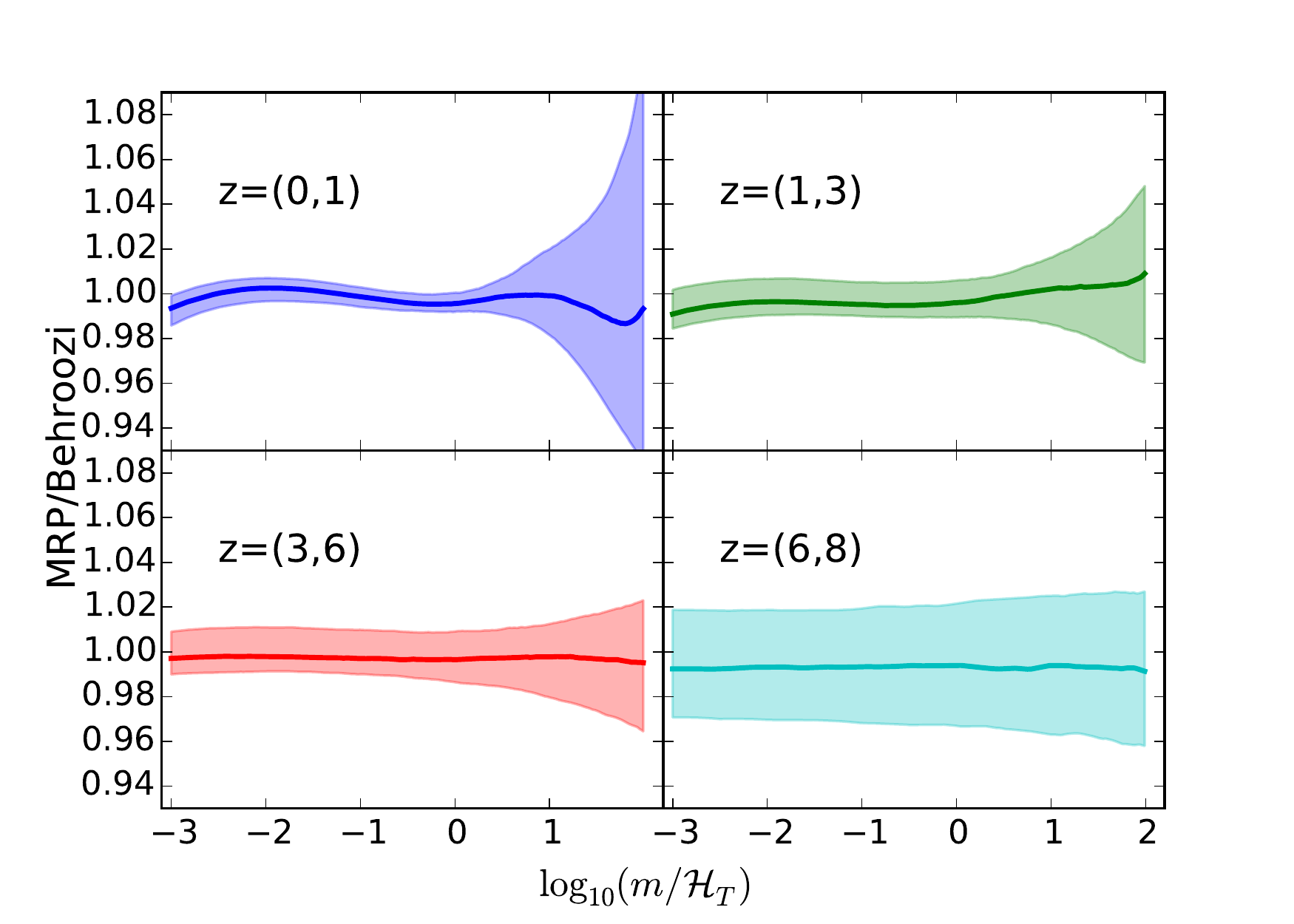}
\caption[Estimated uncertainty in parameterisations for the physical dependence of the MRP parameters]{Central 68\% regions of the relative error for the 3D parameterisations of the MRP parameters compared to the fits of B13. The $x$-axis is the mass normalised to $\mathcal{H}_T$, in which each sample has the same range. Each panel indicates a different redshift range, as marked. The solid line is the median value of each subsample. The uncertainty in the models sub-ten-percent over the entire redshift and mass range.}
\label{mrp:fig:3d_models}
\end{figure}

Figure \ref{mrp:fig:3d_models} shows the 68\% region of relative uncertainties in the models, over the distribution of physical parameters we have employed. The MRP model attained by simply using the formulas from Table \ref{mrp:tab:3d} produces HMFs within 5\% of B13 for almost the entire mass range between $0\leq z \leq 1$, but at high redshifts can deviate by up to $\sim10$\%. Given that uncertainties within the B13 HMF itself are of order 5-10\%, in addition to uncertainties due to cosmology, choice of fitting function and halo finder, this is accurate enough to trust our general results.

\section{The Stellar-Mass Halo-Mass Relation}
\label{mrp:sec:smhm}

The stellar-mass halo-mass (SMHM) relation connects galaxies to halos by relating the typical stellar mass content of a halo to its total mass. 
Determining the form of this relation has been the subject of numerous studies \citep[eg.][]{Moster2010,Behroozi2010,Mutch2013,Durkalec2015}, and is important for quantifying the effects of physical processes such as Supernova and AGN heating.
Given that the fundamental motivation for defining the MRP form is to enable simpler connections between galaxies and halos, in this section we construct a direct formula which ties together the MRP and a double-Schechter parameterisation of the galaxy stellar mass function (GSMF).
In this simplistic setting, we can derive accurate formulae involving both MRP and Schechter parameters which serve to illustrate how a direct parameterisation of the HMF can be conceptually useful.

One simple method of estimating the SMHM relation is to use observed GSMFs, along with theoretical HMFs. 
If using a subhalo-calibrated HMF, there must be at most one galaxy per halo. 
It is generally assumed also that each (sub)halo contains a galaxy, so that each halo contains exactly one galaxy. 
If we also assume that the relationship between a halo mass and its corresponding stellar mass is both deterministic and monotonic (an unrealistic assumption, though studies suggest that the scatter in the SMHM relation is small and constant over mass), then we may derive a relationship between $\ms$ and $m_h$ by
\begin{equation}
n_g(>\ms) = n_h(>m_h).
\end{equation}

Typically, $n_h(>m_h)$ is derived by the PS formalism, and thus can be treated somewhat as a black box in this equation. Though the equation can be solved numerically for $\ms(m_h)$ given any observed GSMF, it is convenient to be able to solve it, at least approximately, analytically in terms of the parameters of the underlying GSMF and the MRP.

The GSMF is commonly accurately parameterized as a double-Schechter function \citep{Baldry2012}, which has the integral
\begin{equation}
n_g(>m_\star) = \Phi_1 \Gamma\left(\alpha_1 + 1, \frac{M_\star }{\Ms}\right)+\Phi_2 \Gamma\left(\alpha_2 + 1, \frac{M_\star }{\Ms}\right),
\end{equation}
where without loss of generality we will assume that $\alpha_2 < \alpha_1$. If we let the HMF be in the form of the MRP, we then arrive at the expression
\begin{equation}
\label{mrp:eq:full-smhm}
\Phi_1 \Gamma\left(\alpha_1 + 1, \frac{M_\star }{\Ms}\right)+\Phi_2 \Gamma\left(\alpha_2 + 1, \frac{M_\star }{\Ms}\right) = A \Hs \Gamma(z_h, x_h).
\end{equation}
Unfortunately, this is analytically unsolvable. However, it is easy to solve by root-finding, as it has a single root and simple derivatives. Thus a numerical procedure is able to produce the empirical SMHM relation. 

\begin{figure}
	\centering
	\includegraphics[width=\linewidth]{\mypref 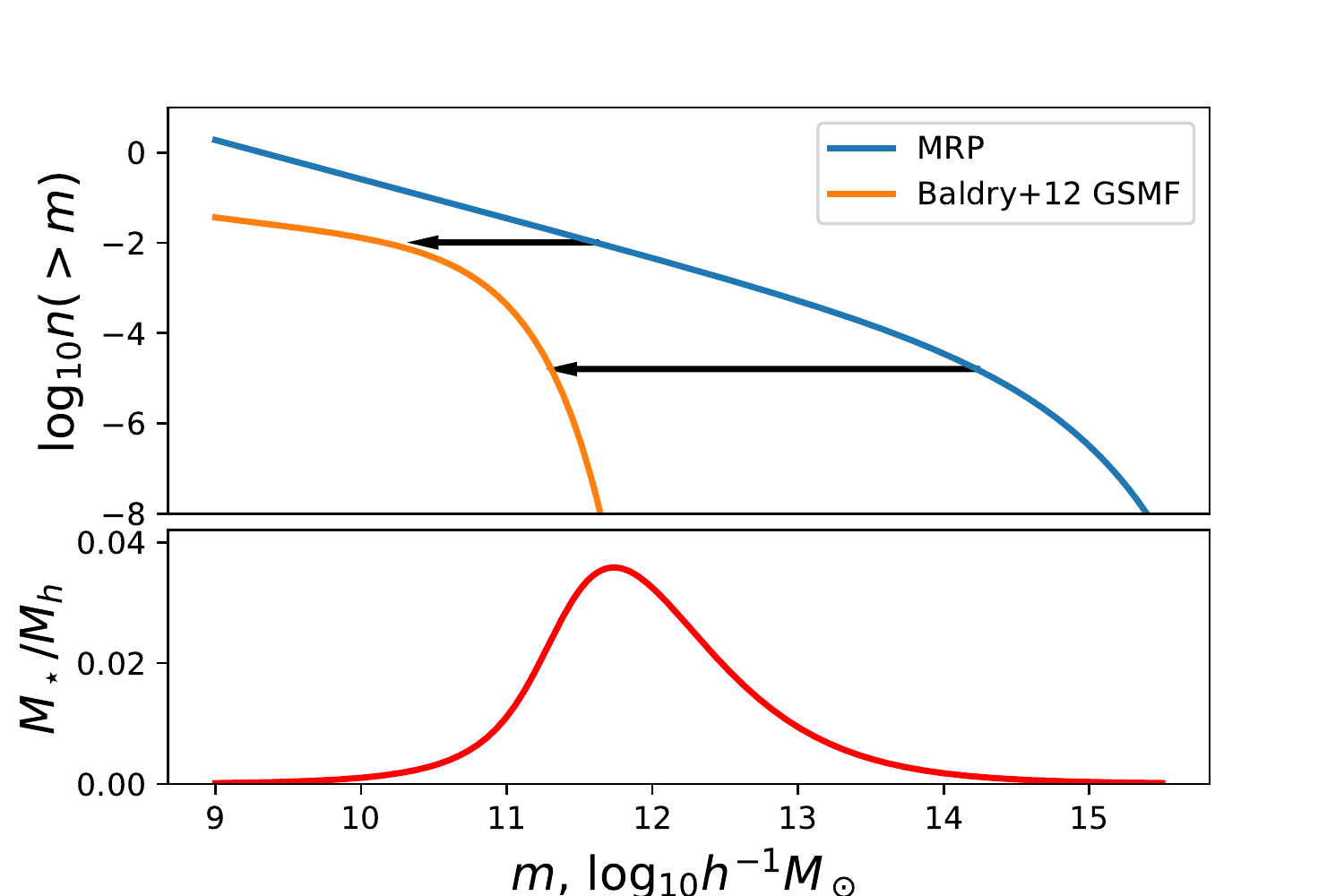}
	\caption[Schematic of stellar mass function]{\textit{Upper panel:} the cumulative MRP corresponding to T08 at z=0 in blue, and the cumulative double-Schechter parameterisation of the z=0 GSMF from \protect\citet{Baldry2012} in orange. Identifying the SMHM relation equates to computing the transform of halo mass to stellar mass, as shown by the arrows. \textit{Lower panel:} The numerically derived fraction $M_\star/M_h$ from the functions in the top panel. There is a strong peak efficiency at $M_h \sim 10^{11.8} h^{-1}M_\odot$.}
	\label{mrp:fig:smf_v_hmf}
\end{figure}

This procedure is illustrated in figure \ref{mrp:fig:smf_v_hmf}, where the cumulative halo mass function is shown in blue, and the cumulative stellar mass function in orange. Given equation \ref{mrp:eq:full-smhm}, the procedure to calculate the ratio $M_\star/M_h$ is to identify the ratio of the curves in the horizontal direction. This can be simply performed by splines, or more robustly by root-finding, to find the red curve in the lower panel. This curve displays the efficiency of producing stellar mass as a function of halo mass, and exhibits a characteristic peak around $10^{12} h^{-1}M_\odot$. 

\begin{deluxetable}{lcl}
	\tabletypesize{\footnotesize} 
	\tablecolumns{3} 
	\tablewidth{0pt} 
	\tablecaption{Summary of SMHM relations in the literature}
	\tablehead{\colhead{Ref.} & \colhead{\# par.} & Formula} 
	\startdata
	M13 & 3 & $\displaystyle \epsilon \exp \left(-\left[\frac{m_h -m_{peak}}{\sigma}\right]^2\right)$ \\
	M10 & 4 & $\displaystyle 2 f_0 \left[\left(\frac{M_h}{M_1}\right)^{-\beta} + \left(\frac{M_h}{M_1}\right)^{\gamma}\right]^{-1}$ \\
	B10 & 5 & $\displaystyle \frac{\ms}{M_0 \left(\frac{M_\star}{M_1}\right)^\beta 10^{\left(\frac{M_\star}{M_1}\right)^\delta/\left(1+\left(\frac{M_\star}{M_1}\right)^{-\gamma}\right)-1/2}}$ 
	\enddata
	\tablecomments{Parameterisations of several SMHM relations from the literature. Note that B10 is in terms of $M_\star$ rather than $M_h$, and so must be numerically inverted (except in limiting cases where it can be analytically inverted). Note also that parameters may have different meanings across parameterisations.}
	\label{mrp:tab:smhm-models}
\end{deluxetable}


\begin{figure}
	\centering
	\includegraphics[width=\linewidth]{\mypref 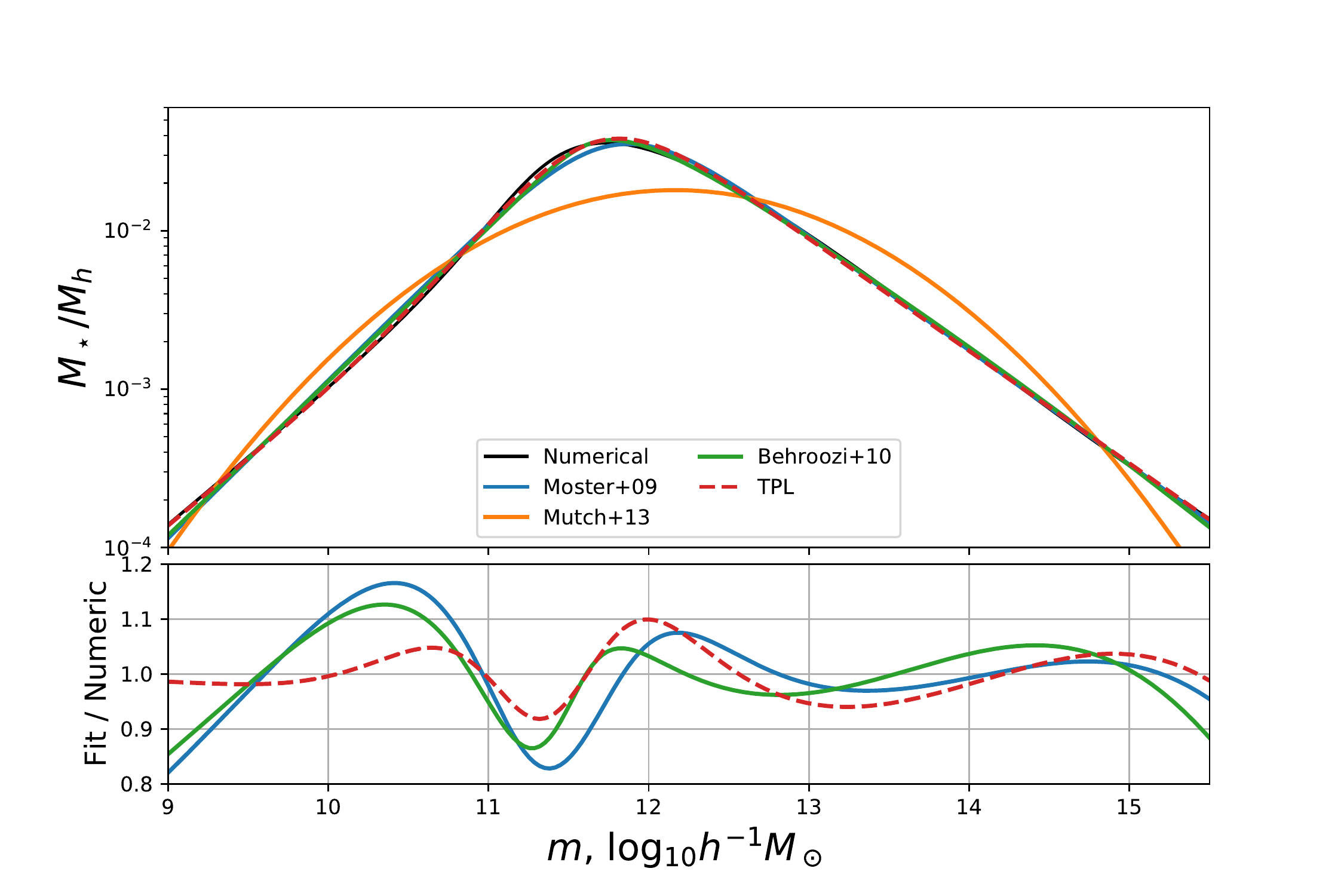}
	\caption[Numerically calculated SMHM with 3 parameterisations from the literature]{The numerically calculated SMHM relation (black, largely hidden by thick red line except just below the turning point), with three best-fit models from the literature. The lower panel shows their ratios. All fits are in log-log space.}
	\label{mrp:fig:smhm}
\end{figure}


It is convenient to have a closed form for the SMHM relation (or equivalently $f=M_\star/M_h$). 
Several have been proposed in the literature, each of which lies at a different point on the spectrum of simplicity versus flexibility.
We list three such parameterisations in Table \ref{mrp:tab:smhm-models} (\citet{Mutch2013}, \citet{Moster2010} and \citet{Behroozi2010} respectively), and show an instance of each in Figure \ref{mrp:fig:smhm}, where each is fit against the numerically-calculated black curve (we note that this curve is calculated under all the assumptions of strict halo abundance matching as described above, and the goodness-of-fit of these curves is not an indication of their accuracy to observed data or their general usefulness).
In this section, we add our own parameterisation, based on solving Eq. \ref{mrp:eq:full-smhm}.

Though we cannot solve Eq. \ref{mrp:eq:full-smhm} for all $m_h$, we do well to identify its behaviour in the limits. 
As $x \rightarrow 0$, we have the identity $\Gamma(z,x) \rightarrow -x^z/z$, for $z<0$ (this can be simply derived from the recursion relation, Eq. \ref{mrp:eq:recurrance}). The speed of convergence of this limit depends heavily on the value of $z$ (the more negative, the faster the convergence). 

In this case, at small mass, the $\alpha_2$ term dominates the $\alpha_1$ term (which generally has positive shape parameter), so we can simply write
\begin{equation}
-\frac{\Phi_2\left(\frac{M_\star}{\Ms}\right)^{(\alpha_2 + 1)} }{\alpha_2 + 1} = -\frac{A \Hs x_h^{z_h}}{z_h},
\end{equation}
so that we find that the ratio is a power law:
\begin{align}
\ms - m_h &= \log_{10}K +  p m_h, \\
p &= \frac{\alpha_h-\alpha_2}{\alpha_2+1} \\
\log_{10}K &= \ms  - \frac{\alpha_h}{\alpha_2+1}\hs + \frac{1}{\alpha_2+1} \log_{10} \left(\frac{A (\alpha_2+1)}{\Phi_2 z_h}\right),
\end{align}
Thus we favour a parameterisation that is a power-law in the lower limit, such as M10 or B10 in Table \ref{mrp:tab:smhm-models}.
Specifically, M10 is equivalent in the low-mass limit if $\beta = p$ and $2 f_0/M_1^p = K$, while B10 requires $\beta = 1/(p+1)$ and $M_1 (\sqrt{10}/M_0)^{1/\beta} = K$.


Unfortunately, while the high-mass limit affords an analytic asymptotic approximation, we are more interested in scales $\ms/\Ms \lesssim 10$ (higher masses are almost non-existent). 
We have not been able to derive an approximation in this regime. 
However, the power-law behaviour exemplified M10 and B10 seems sufficient. 

The most severe deficiency remaining is the position of the turning point. 
A simple way to change the position of the turning point without affecting the behaviour in the limits, is to use an extension of M10:
\begin{equation}
\frac{M_\star}{M_h} = \frac{w\left(\frac{M_h}{M_1}\right)^{-\delta} + 2 f_0}{\left(\frac{M_h}{M_1}\right)^{-\beta} + w\left(\frac{M_h}{M_1}\right)^{\gamma} + k},
\end{equation}
where $w$ controls the peak position, $k$ is able to correct the amplitude of the high-mass power-law, and $\delta$ adds the flexibility needed to induce an upturn left of the turning point. 
This extension obeys the same relations as M10 in terms of the low-mass approximation, but has the difference that the high-mass power law has the slope $\delta-\gamma$. 

Using the two fixed parameters from these relations, the resultant best-fit curve is plotted in magenta in figure \ref{mrp:fig:smhm} (``TPL", for `triple power-law'). We find much better agreement in terms of the turning point position, though the precise shape of the maximum is clearly still not reproduced. 
With the restrictions enforced, the model has the same number of parameters as the B10 form.

\section{An Application: Survey Design}
\label{mrp:sec:survey}
In this section we use many of the results of the previous sections in a survey design application. 
In particular, we determine a semi-realistic model for the observations of a galaxy survey which determines group masses dynamically. 
While it is clear that this model ignores some of the complications associated with a realistic galaxy survey, it nonetheless attempts to capture some of this complexity, and show how the MRP in conjunction with the robust Bayesian parameter estimation of \S\ref{mrp:sec:fitting} can be used in realistic situations.

Our key question in this section will be: for a galaxy survey within a reasonably thin redshift slice with a fixed total number of observed galaxies, are the MRP parameters better fit by (i) increasing the sky area, (ii) probing fainter objects or (iii) making more aggressive quality cuts?

Our broad approach is to first define a model for the underlying distribution of halo masses, their observables, and their measurement uncertainties. 
This forward model in turn provides the relevant functions for use in Eq. \ref{mrp:eq:basic_lnl} to determine the likelihood of a given set of data.
Instead of producing synthetic data and generating parameter estimates for it, we will focus on the expected covariance of parameters under our data model (cf. \S\ref{mrp:sec:expected_cov}).

\subsection{Data Model}
We will here exclusively consider an observation within a solid angle $\Omega$, with no angular selection function, and within a small redshift slice, $z\in (z_0,z_1)$. 
Specifically, the width of the redshift slice is assumed to admit negligible evolution in the HMF parameters $\thp$. 
The (mean) redshift of the observation is thus taken to be 
\begin{equation}
z = 0.75 \frac{z_1^4-z_0^4}{z_1^3-z_0^3}
\end{equation}
and the total volume is
\begin{equation}
 V_0 = \frac{\Omega}{4\pi {\rm sr}}  [V_c(z_1) - V_c(z_0)],
\end{equation}
where $V_c$ is the comoving volume out to a given redshift over the whole sky \citep{Hogg1999}. Furthermore, we specify the depth of the survey as a limiting apparent magnitude, $m_0$.

We can simply generate the limiting absolute magnitude using the following relation:
\begin{equation}
M_0 = m_0 - (5\log_{10}D_L(z) + K(z) + 25),
\end{equation}
where $K(z)$ is the $K$-correction \citep{Hogg2002} and $D_L$ is the luminosity distance \citep{Hogg1999}.

Now let the luminosity function be denoted $\phi_L(M)$, and for it we employ a single-Schechter function (with parameters given by \cite{Loveday2012}). In addition, let the stellar mass function (GSMF) be denoted $\phi_\star(m_\star)$, which will be a double-schechter function (with parameters given by \cite{Baldry2012}). Using abundance matching, we can specify the limiting stellar mass by equating
\begin{equation}
\int^{m_{\star,0}} \phi_\star(m_\star) dx_\star = \int^{M_0} \phi_L(M) dM,
\end{equation}
and solving for $m_{\star,0}$ (this would typically require numerical splines or root-finding).

Let the underlying halo mass function be $g(m|\thp)$. Furthermore, let the function $f(m|\thp)$ define the mean ratio of stellar-to-halo mass, $M_\star/M$ for a given halo mass (i.e. the SMHM relation). 
We will use the best-fit TPL model shown in Fig. \ref{mrp:fig:smhm}.
Then the average total stellar mass content in a halo of mass $m$ is
\begin{equation}
\label{eq:mstar_to_m}
m_\star^T = m + \log_{10} f(m|\theta).
\end{equation}
This mass is, on average, comprised of a number of galaxies drawn from the GSMF. Note here we are cutting a corner -- the SMHM relation is typically defined using abundance matching between the GSMF and the HMF. That is, the SMHM itself assumes a single galaxy of a given mass in any halo. We continue to assume this for the SMHM relation definition, but change this assumption at this step to that of a number of galaxies constituting the given mass in any halo. We do not worry about conditional mass functions here, but identically draw galaxies from the universal GSMF within each halo. 

What we desire is the distribution of galaxy mass counts for which the sum totals a given mass, i.e. the distribution of $N$ such that
\begin{equation}
\sum_i^N 10^{X_i} = 10^{m_\star^T},
\end{equation}
where $X_i \sim \phi_\star(m_\star)$. The solution to this problem is provided by the hitting time theorem \citep{Hofstad2008}, however the solution is rather intractable for our distributions. 

Instead, we find an approximate answer by first generating an overall normalisation of the GSMF required to generate the mass $m_\star^T$:
\begin{equation}
D = \frac{m_\star^T}{\int^{m_\star^T} M_\star \phi_\star(m_\star) dm_\star}.
\end{equation}
We then use this normalisation to determine the average number of galaxies that make up the mass in the halo:
\begin{equation}
\bar{n}(m_\star^T) = D\int_{m_{\star,0}}^{m_\star^T} \phi_\star(m_\star) dm_\star.
\end{equation}

What this model ignores is the enforced correlation between masses due to the necessity of summing to $x_\star^T$, and the discrete nature of the distribution. We expect that these effects will be second-order, and for our purposes will be negligible. We make up for the latter in some respects by assuming that the occupation of the halo is Poisson distributed. 
We can generate the total fraction of observed halo masses, given some threshold for mass estimation, $n_{\rm min}$, by calculating the CDF of the Poisson distribution at $n_{\rm min}$:
\begin{equation}
f_{\rm obs}(m_\star^T) = 1 - \frac{\Gamma(n_{\rm min} + 1, \bar{n}(m_\star^T))}{n_{\rm min}!}.
\end{equation}
Furthermore, we use the empirical relation of \citep{Robotham2011} to specify mass uncertainties on the dynamically-estimated masses:
\begin{equation}
\sigma_i = {\rm max}\left(0.02, 1 - 0.43\log_{10} N_{{\rm FoF},i} \right),
\end{equation}
with 
\begin{equation}
N_{\rm FoF} \sim {\rm Poiss}(\bar{n}(m_\star^T)). 
\end{equation}

Thus we arrive at the following forward model:
\begin{subequations}
\begin{align}
V(m|\thp) &= f_{\rm obs}(m_\star^T(m|\thp)) V_0 \\
m &\sim V(m|\thp)\phi(m|\thp) \\
N_{\rm FoF} &\sim {\rm Poiss}(\bar{n}(m_\star^T(m|\thp))_{n_{\rm min}} \\
\sigma &= {\rm max}\left(0.02, 1 - 0.43\log_{10} N_{{\rm FoF},i} \right) \\
m' &\sim \mathcal{N}(m,\sigma)
\end{align}
\end{subequations}

We note that in this forward model we have avoided the fact that the mass-observable is in fact the velocity dispersion of the galaxies, other than incorporating an empirical relation for the uncertainty of this proxy. 
We note also that two ``observables" are present in the model, both the mass (via the velocity proxy) and the number of galaxies in the halo. 
There is a temptation to neglect the latter in favour of directly using the derived uncertainties, $\sigma$, however this proves to be an unreliable method as it ignores vital information.

Our application here will be to determine the constraints on $\thp$ for a range of survey parameters, $\Omega$, $m_0$ and $n_{\rm min}$.

\subsection{Likelihood and Covariance}
The basic task is to specify $n(\vec{x}'|\thp)$, the expected number density of observables, $\vec{x}' = (m',N_{\rm FoF})$. 
This clearly requires the three sampling statements from the forward model above, and is merely integrated over underlying mass $m$:
\begin{align}
	n(\vec{x}'|\thp) = &\int dm V(m|\theta) g(m|\vec{\theta}) e^{-(m-m')^2/2\sigma^2(m)} \nonumber \\
	& \times \frac{\bar{n}(m)^{N_{\rm FoF}} e^{-\bar{n}(m)}}{N_{\rm FoF}!}\frac{1}{1-\sum_{j=0}^{n_{\rm min}} \bar{n}(m)^j e^{-\bar{n}(m)}/j!}.
\end{align}
The second line of this equation is the truncated Poisson distribution. For clarity, we express the average observed halo occupancy as $\bar{n}(m)$, while noting that the dependency on the underlying mass arises explicitly through Eq. \ref{eq:mstar_to_m}, and that through this equation it also implicitly depends on $\thp$. 

With this expression, the likelihood of a given set of data, according to Eq. \ref{mrp:eq:basic_lnl}, is
\begin{equation}
\ln L(\vec{\theta}) = - n_\theta  +  \sum_i \ln n(m'_i, N_{\rm FoF,i}|\thp),
\end{equation}
with $n_\theta$ the expected number of observations, $n_\theta \equiv \int Vg dm$. 

The expected likelihood is gained by converting the sum over observations into a sum over infinitesimal bins in the observable space -- i.e. conversion into a density-weighted integral (cf. \S\ref{mrp:sec:expected_cov}).
In our case, the observable space is 2-dimensional, and one of the dimensions is discrete (i.e $N_{\rm FoF}$). 
Thus we arrive at an expected likelihood of
\begin{equation}
	\ln L = - n_\theta + \int dm' \sum_{j=n_{\rm min}} n'(m',j) \ln n(m',j).
\end{equation}
We may also use the same reasoning to derive the expected hessian, using Eq. \ref{mrp:eq:expected_hessian}.

\subsection{Results}
We define the parameters of the data model such that observations are between $z \in (0.07, 0.13)$, and use an input $\thp$ gained from the formulae of table \ref{mrp:tab:3d} at the mean redshift.
We then calculate the expected covariance at a range of values of $m_0$ and $n_{\rm min}$, with the sky area $\Omega$ varied so that the total number of observed galaxies (including those in systems with fewer than $n_{\rm min}$ galaxies) is $5\times10^4$. 

\begin{figure*}
\centering
\includegraphics[width=0.9\linewidth]{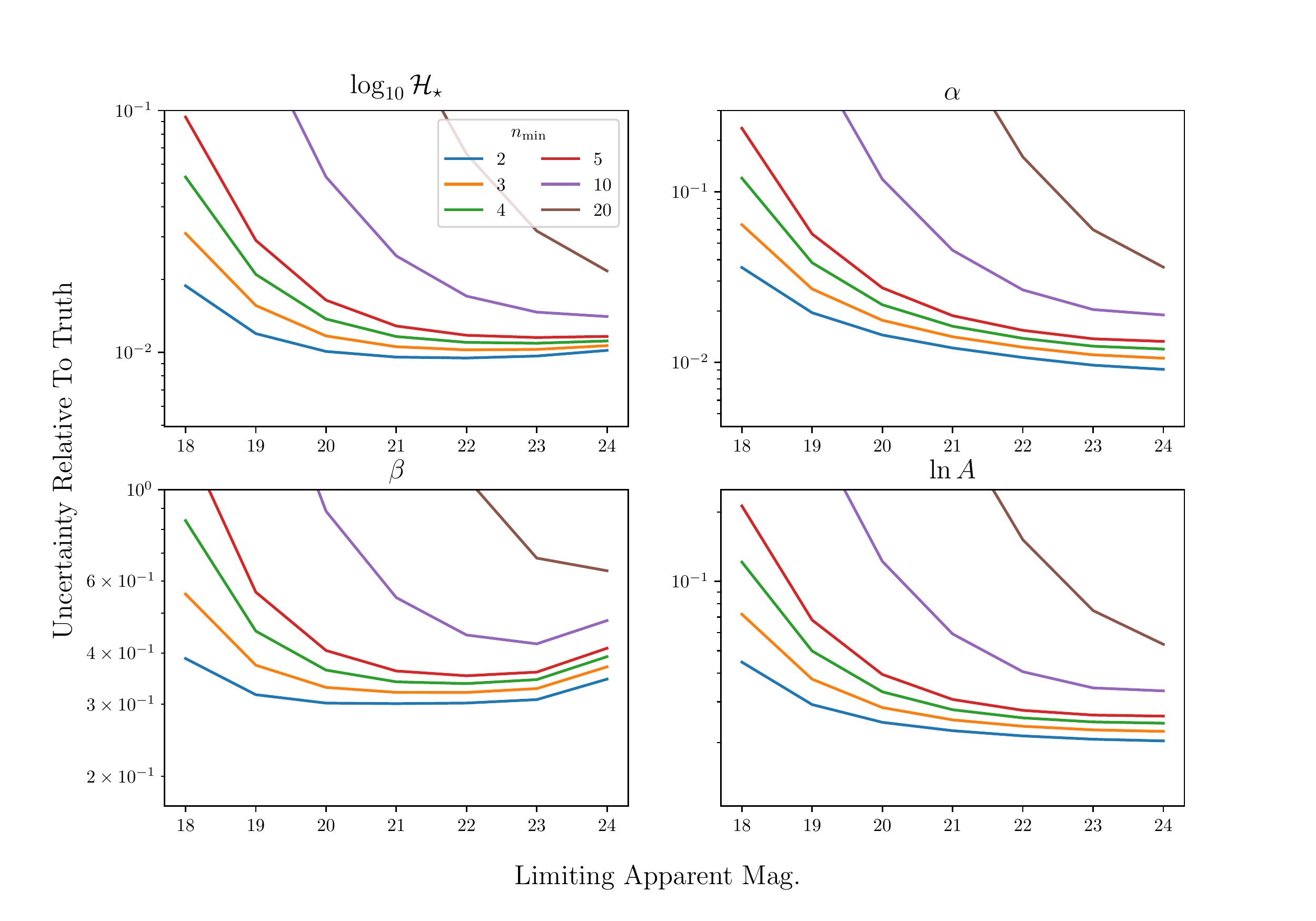}
\caption[Marginal variance of MRP parameters in complex data model]{The marginal variance of the MRP parameters. The $x$-axis is the limiting apparent magnitude, $m_0$, while the $y$-axis is the marginalised standard deviation of the parameter as a fraction of its true value. Different colours represent different threshold occupations.}
\label{fig:application_fig}
\end{figure*}

Figure \ref{fig:application_fig} shows the results of this analysis.
In particular, the depth of the observation increases to the right in each panel, while the occupation threshold (i.e. the fewest number of galaxies a group must contain for inclusion in the sample) is denoted by the colour of the curve. 

We first consider the global properties of the uncertainty. 
We see that other than $\beta$, the parameters are typically fit to better than 10\%, approaching 1\% for $\hs$ and $\alpha$. 
The poorer fit of $\beta$ is expected as it is solely probed by the scant high-mass haloes. 

As a function of $n_{\rm min}$ there is perhaps surprising unanimity on the result that a lower truncation yields a tighter posterior. 
This means that the extra raw statistics brought about by increased sample size outweighs the relative degradation in sample quality that a lowering in truncation brings. 

Finally, as a function of depth, there are two groups. Firstly, $\alpha$ and $\ln A$ monotonically increase in precision as the depth increases (though there is a clear flattening of this relationship above $m_0 \sim 20$). 
Conversely, $\hs$ and $\beta$ initially increase their precision, but subsequently turn around and yield decreased precision for the deepest surveys. 
This can be intuitively understood. 
Initially, as the depth increases, the average quality of the data increases for the same average number of sources.
However, at the same time, the range of masses probed is increased.
As more and more low-mass haloes are added to the sample, they begin to dominate the fit, and the high-mass haloes to which $\beta$ and $\hs$ are sensitive become relatively insignificant. 
At some point, around $m_0 \sim 20-23$, the latter effect wins out, and the precision of the fits deteriorates.

This would suggest that an optimal strategy for determining all HMF parameters is to observe down to $m_0 \sim 21$ and use all observed group masses.
While this analysis is rather simplistic, it is offered as an example application to show the potential power of more complex forecasting in future survey designs.

\section{Conclusion}
\label{mrp:sec:conc}
We have introduced a new analytic formula to describe the halo mass function over a broad range of scales, based loosely on the existing PS formalism, and closely related to the popular Schechter functional form. 
Our form is motivated by the desire to concisely and intuitively measure the HMF down to group-scale haloes, and provide simple comparisons to galaxy statistics such as the stellar mass function and luminosity function.

We have shown that the MRP is an accurate model, within the uncertainties inherent in the mass function due to halo finders, halo definition and simulation resolution. 

To exemplify the utility of our functional form, we determined simple fitting functions for the dependence of the MRP parameters on the physical parameters $z$, $\Omega_m$ and $\sigma_8$, showing that each of the parameters primarily has a negative correlation with redshift. Furthermore, we showed that it is possible to analytically motivate more precise parameterisations of the stellar-mass-halo-mass relation by using the known properties of the MRP in conjunction with the common double-Schechter form for the stellar mass function.

We presented in some detail the machinery required to use the MRP form to fit to halo catalogues -- both simulated and observed. 
Importantly, we highlighted the Bayesian methods required to incorporate arbitrary measurement uncertainties into the analysis.
We also presented novel results concerning the \textit{expected} likelihood and covariance within a given data model.

Ultimately, we used these results to perform an example application, in which we determined the expected errors on fitted MRP parameters in a complex data model including mass-dependent selection effects and dynamical mass uncertainties. 
We showed that in such a model, with a constant number of observed galaxies, the optimal depth of the survey is an apparent magnitude of approximately 21. 

We have highlighted several areas in which the work presented here might be extended throughout the course of this paper. 
In particular, investigations into alternate parameterisations which have better covariance properties may be beneficial. 
However, the best extensions will be to use the framework presented to fit actual datasets and begin to relate HMF parameters to those of the galaxy population.

\section*{Acknowledgments}
SGM would like to thank Ian McCarthy and Ishiyama Tomoaki for access to and help with simulation data.
ASGR acknowledges support of a UWA postdoctoral research fellowship, as well as an Australian Research Council (ARC) Discovery Project (DP) grant, DP140100395.
The Centre for All-Sky Astrophysics (CAASTRO) is an Australian Research Council Centre of Excellence, funded by grant CE11E0090. 
This research has made use of NASA's Astrophysics Data System.

\ifx\isEmbedded\undefined
\bibliography{library}
\begin{appendix}
\ifx\isIncluded\undefined
    \documentclass{article}
    \usepackage{natbib}
    \usepackage{graphicx}
    \usepackage{amsmath}
    \usepackage{amssymb}
    \usepackage{hyperref}
    \title{Introduction}

    \newcommand{\mypref}{}
    \begin{document}

\fi

\section{Recurrence Relation}
\label{mrp:app:recurrence}
A problem that arises in the calculation of the incomplete gamma function is that several libraries only implement $\Gamma(z,x)$ for $z>0$, and we often need it for $z<0$. One way around this is to use the recurrence relation
\begin{equation}
	\label{mrp:eq:recurrance}
	\Gamma(z,x) = \frac{\Gamma(z+1,x)-x^z e^{-x}}{z},
\end{equation}

In general, for arbitrary negative $z$, one can use the following extended relation to calculate the incomplete gamma function in terms of only positive shape parameter:
\begin{equation}
	\label{mrp:eq:gen_rec}
	\Gamma(z,x) = \Gamma(z)\left(Q(z+n,x) - x^z e^{-x} \sum_{k=0}^{n-1} \frac{x^k}{\Gamma(z+k+1)}\right),
\end{equation}
where $Q$ is the regularized incomplete gamma, $Q(z,x) = \Gamma(z,x)/\Gamma(z)$. To ensure the use of only positive shape parameters, $n$ should be set to $\lceil -z \rceil$. Note that the accuracy of this formula will decrease for large $|z|$ due to numerical precision, and likewise the evaluation time will increase due to the sum.

\section{Re-parameterisations}
\label{mrp:sec:repar}
There are several re-parameterisations of the GGD to be found in the literature, which could potentially bring better covariance properties than vanilla MRP. In this section, we investigate three of these formulations, along with one that we have devised.

Table \ref{mrp:tab:repars} summarises the parameterisations we will investigate here. There is also at least one form in the literature which claims to have better properties than those listed here \citep{Lawless1982}, but is purely defined for $z>0$, and therefore cannot be used.

The HT formulation has the immediate benefit that two of the parameters have physical meanings: $\nu$ is the power-law slope, and $\mu \equiv \mathcal{H}_T$ is the logarithmic mass mode. We expect this to be a very robust scale in any given data in which it exists (it will not exist when the slope is $\leq-2$), and should be able to be fit almost independently of the other parameters.

In non-heirarchical contexts, we may simply test the expected covariance of any re-parameterisation by converting the identified covariance of the MRP form to any re-parameterisation for which we have the derivatives of the transform. 

Given a new vector of parameters, $\vec{\psi} = (\mu, \nu, \delta)$, and a set of three equations $\thp(\vec{\psi}) = \thp$ relating the parameters, then the Jacobian at any point $\vec{\psi}$ is given by
\begin{equation}
	J_i^\psi = \vec{J}^\theta \cdot \frac{\partial \thp}{\partial \psp_i}.
\end{equation}
We can then calculate the Hessian as
\begin{equation}
	\label{mrp:eq:convert_cov}
	H^\psi_{ij} = \vec{J}^\theta \cdot \frac{\partial \thp}{\partial \psp^i \partial \psp^j} + \frac{\partial\thp}{\partial \psp^j} \cdot \mathbf{H}^\theta \cdot \frac{\partial \thp}{\partial \psp^i}.
\end{equation}

To test the covariance properties of these parameterisations with respect to the MRP form, we generate mock halo catalogues by sampling the MRP at a typical point of $\thp=(14.5,-1.85,0.72)$ (corresponding roughly to P13 best-fit values). We do this for a series of differing truncation masses (from $m_{\mathrm min} = 10$ to 13) and for each, we produce 20 realisations, calculating the hessians at the input point. We convert these into the new parameterisation space using Eq. \ref{mrp:eq:convert_cov}, and then convert the result into covariance and correlation matrices. 

\begin{figure}
\centering
\includegraphics[width=0.75\linewidth]{\mypref 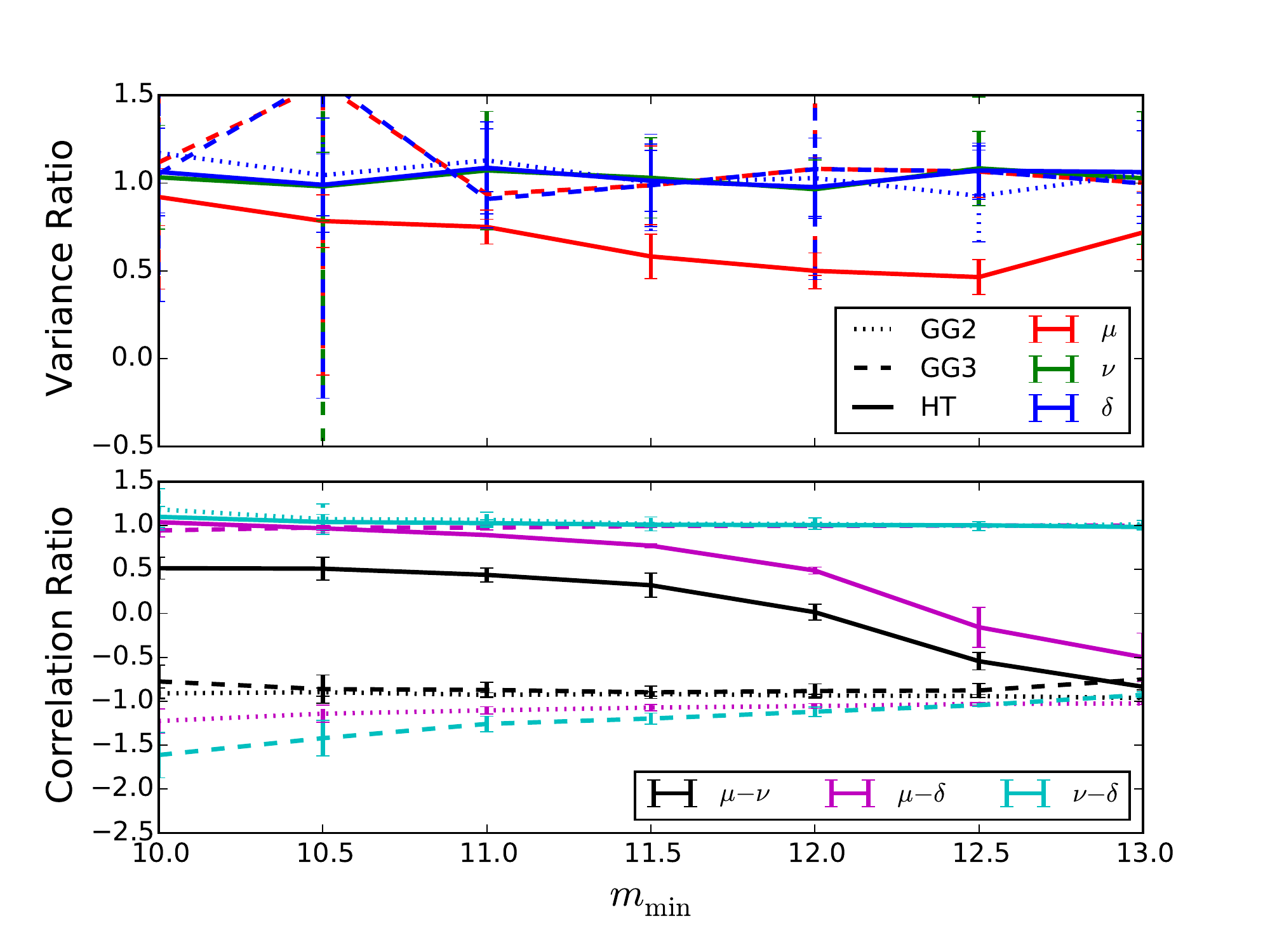}
\caption[MRP re-parameterisations]{Properties of three reparameterisations of the MRP compared to the form used in this paper. $x$-axis represents truncation mass. All dotted (dashed, solid) lines refer to GG2 (GG3, HT) respectively. The colours represent different quantites -- the red (green,blue) line in the top-panel is the ratio of the standard deviation of $\mu$ ($\nu$, $\delta$) to $\hs$ ($\alpha$, $\beta$) at the solution (where the standard deviation is relative to the value at the solution). The black (magenta, cyan) line in the lower panel is the ratio of the correlation coefficient between $\mu-\nu$ ($\mu-\delta$, $\nu-\delta$) to the corresponding MRP correlations. All ratios are taken as the means of 20 realisations from ideal mock catalogues.}
\label{mrp:fig:repars}
\end{figure}

Figure \ref{mrp:fig:repars} shows the mean results of this analysis, with errorbars. Each of the linestyles represents different parameterisations, while the colours represent different parameter combinations. The top panel shows the ratios of variances in the new parameterisation to MRP. The variances are all normalised to their parameter values. This indicates which parameterisations are more effective at net constraints on the parameters. We find that, as expected, $\delta \sim \beta$ is not improved for any of the alternate forms. On the other hand, $\nu$ is estimated similarly for HT, but is outside the range of the plot for both GG2 and GG3, and so has poorer precision. However, $\mu$ varies widely between the new forms. For GG2, it has a far poorer relative deviation, while for GG3, as expected, it remains consistent with unity. For HT, it is significantly more precise -- for some truncation masses it has double the precision. 

The double-parameter quantities in the lower panel are the ratios of the correlation coefficients between the new parameterisations and the MRP. While the $\mu-\nu$ correlation swaps sign for GG2 and GG3, it perhaps surprisingly remains consistent with a parity of magnitude. Conversely, for HT, the correlation is rather dependent on $m_{\mathrm min}$, but is always significantly lower than MRP, as expected. We find much the same result for $\mu-\delta$, though HT only outperforms MRP for high $m_{\mathrm min}$. Finally, $\nu-\delta$ is relatively unchanged in all parameterisations except for GG3, in which it performs poorly at low $m_{\mathrm min}$. 

We conclude that neither GG2 or GG3 are useful alternatives for the MRP in the context of typical values for the HMF, having no beneficial properties. Conversely, HT is either beneficial or benign in every aspect, and provides an interesting re-formulation for future studies. Its major potential drawback is that it is undefined for $\nu \leq 2$, which may be transgressed in some rare cases for certain values of $m_{\mathrm min}$. 

An interesting corollary is that the same situation applies to the popular Schechter function. In that case, $\beta \equiv 1$, so that $\mathcal{H}_T = \mathcal{H}_s (\alpha+2)$. Since $\alpha \approx -1$, $\mathcal{H}_s$ has often been directly associated with the turnover. However, this is not strictly true, and for common measurements of $\alpha$ between -1.4 and -1.1, there may be a corresponding benefit in using the true turnover scale. 

 \begin{deluxetable}{lccc}
 	\tabletypesize{\footnotesize} 
 	\tablecolumns{4}
 	\tablewidth{0pt}
 	\tablecaption{}
 	\tablehead{\colhead{Ref} &  \colhead{$\hs$} & \colhead{$\alpha$} & \colhead{$\beta$}} \\
 	\startdata
 	GG2 \citep{LagosAlvarez2011} & $\mu^{-1/\delta}$ & $\nu-1$ & $\delta$ \\
 	GG3 \citep{LagosAlvarez2011} & $\mu$ & $\delta\nu-1$ & $\delta$ \\
 	HT & $\mu\left(\dfrac{\delta}{\nu+2}\right)^{1/\delta}$ & $\nu$  & $\delta$
 	\enddata
 	\tablecomments{Alternate parameterisations found in the literature, or presented here. Columns indicate the transformation from the new parameters $(\mu,\nu,\delta)$ to the standard $(\Hs,\alpha,\beta)$. We note that in the HT formula, $\mu$ is equivalent to the logarithmic mass mode $\mathcal{H}_T$. }
 	\label{mrp:tab:repars}
 \end{deluxetable}

\section{Cosmic Variance and Other Uncertainties}
\label{mrp:app:cosmic_variance}
Throughout this paper we have used a likelihood which is exact when the model prediction of the counts per infinitesimal bin is known, and the distribution from which the bin counts are taken is Poisson. 
In practice, the model prediction itself contains uncertainty due to cosmic variance, or even due to an uncertain selection function. 
In this section, we provide two methods of dealing with such uncertainty, both of which unfortunately require some form of \textit{ad hoc} binning. 

In general, the cosmic variance will have a \textit{covariant} effect on the counts -- bins with similar masses will tend to be correlated (while the Poisson uncertainty remains independent). 
One can account for this by employing an arbitrarily fine mass binning such that we have a vector $\vec{X}$ of counts, of length $N_{\rm bins}$.
The first method of incorporating cosmic variance relies on integrating over the extra uncertainty.
The expected counts per bin is given by $\vec{\lambda} = \vec{V}\vec{\phi}$, and we have elsewhere assumed that for an infinitesimal bin, the number counts are drawn as $X_i \sim {\rm Poiss}(\lambda_i)$.
However, with cosmic variance, the value of $\vec{\lambda}$ contains uncertainty. 
Let $f_{\rm cv}(\vec{\lambda}'|\vec{\lambda})$ be the probability distribution function of values of $\vec{\lambda}$, given the prediction from the model $V\phi$, arising from cosmic variance (for example).
Then the probability mass function of source counts, $X$, is 
\begin{eqnarray}
P(\vec{X}=\vec{k}) = \int d\vec{\lambda}' \prod_i^{N_{\rm bins}} \frac{\lambda_i^{'k_i} e^{-\lambda'_i}}{k_i!} f(\vec{\lambda'}_i|\vec{\lambda}).
\end{eqnarray}
This is a highly multi-dimensional integral which will be typically impractical to solve on any iteration.

The second method is to use a hierarchical Bayesian model. 
This is most easily conceptualised as a generative model with the following steps:
\begin{enumerate}
	\item Choose MRP parameters $\theta$ from their prior distributions.
	\item Calculate $\vec{\lambda}$ for the arbitrary binning employed, which represents the universal background mass function.
	\item Generate a realisation of $\vec{\lambda}'$ for the sample at hand (i.e. $N_{\rm bins}$ new parameters), given the cosmic variance distribution $f_{\rm cv}(\vec{\lambda}'|\vec{\lambda})$.
	\item Generate true masses using a Poisson distribution with mean $\lambda'_i$ in each bin.
	\item Scatter true masses according to a measurement uncertainty distribution.
\end{enumerate}
Such a model can be constrained by using this process on every iteration.
That is, we employ $N_{\rm bins}$ new unknown parameters describing $\vec{\lambda}'$, which are drawn on every iteration. 
Equation \ref{mrp:eq:basic_lnl} is then evaluated (necessarily numerically) using $\lambda'(m)$ rather than $\lambda(m)$.
While this method involves an \textit{ad hoc} binning scheme, it can in principle be tested for convergence by increasing the bin resolution (with the caveat that doing so will require a greater computation time due to the extra parameters it must constrain).

\section{Parameter Estimation with MRP}
\label{mrp:app:jac_hess}
In \S\ref{mrp:sec:fitting} we introduced a robust likelihood for a given choice of parameters, $\thp$, and a sample of data, to be used for parameter estimation (cf. Eq. \ref{mrp:eq:basic_lnl}). 
In addition, we specified the expected covariance at the MLE given an input model (cf. Eq \ref{mrp:eq:expected_hessian}).
In this appendix, we supply additional details to these likelihoods to aid in practical implementations, and we also add details of a different kind of fit, in which the MRP functional form is directly fit to another model using least-squares minimization in log-log space. 

In particular, we present the Jacobian (here interchangeable with the gradient) and Hessian (i.e. the matrix of second-order derivatives) for these likelihoods in the context of the MRP. These latter quantities are useful for accelerating convergence in downhill-gradient methods, and also enable the estimation of parameter covariances at the solution.

\subsection{Method I: Binned/Curve data}
\label{mrp:app:jac_hess_bin}
The basic method for fitting a model to a curve (or equivalently to binned data) is to use a minimization of $\chi^2$. In this case, due to the massive dynamic range of the data (both in mass and number density), it is preferable to perform the fit in log-space. Specifically, the log-likelihood of the model is given by
\begin{equation}
\label{mrp:eq:ll_binned_1_app}
\ln \mathcal{L}_{\mathrm bin}(\thp) = -\frac{\chi^2}{2} = -\sum_i^N \frac{(\ln g(m_i|\thp)-\ln d_i)^2}{2\sigma_i^2} \equiv -\sum_i^N \frac{\Delta_{\rm data}^2}{2\sigma_i^2},
\end{equation}
where $g$ is defined as in Eq. \ref{mrp:eq:g}, $d_i$ is the value of the $i^{th}$ bin of data, and $\sigma_i$ is the uncertainty in the $i^{th}$ bin of data.

This likelihood may be generalised by allowing an extra constraint to play a role, namely that of the Universal mean density, 
$\bar{\rho_0} = \Omega_m \rho_c$. Briefly, this engenders a likelihood of
\begin{equation}
\label{mrp:eq:ll_binned_4}
\ln \mathcal{L}_{\mathrm bin}(\thp) = -\left(\sum_i^N \frac{\Delta_\mathrm{data,i}^2}{2\sigma_i^2} + \frac{\Delta^2_\rho}{2\sigma_\rho^2}\right),
\end{equation}
where
\begin{equation}
    \label{mrp:eq:delta_rho}
    \Delta_\rho = \ln A - \ln \bar{\rho}_0 + \ln k(\thp_3) \\
\end{equation}
and $k$ is defined (cf. Eq. \ref{mrp:eq:k}) as
\begin{equation}
    k(\thp_3) = \Hs^2 \Gamma(z_1).
\end{equation}

In Eq. \ref{mrp:eq:ll_binned_4}, $\sigma_\rho$ controls the tightness of the constraint. A value of infinity provides no constraint, in which case the likelihood is equivalent to Eq. \ref{mrp:eq:ll_binned_1_app}. 
Conversely, a value of zero requires that the parameters be chosen to exactly reproduce $\bar{\rho}_0$. In this case, the estimated parameters are $\thp_3 \equiv (\hs,\alpha,\beta)$, and Eq. \ref{mrp:eq:delta_rho} is used to determine $\ln A$. 

We provide this constraint purely for theoretical interest -- using it in practice could lead to systematic biases since the behaviour of the HMF below $m_{\rm min}$ is not well determined.

\subsubsection{Jacobians and Hessians}

We here only derive the Jacobian and Hessian in the case that $\sigma_\rho \neq 0$.
In this case, the Jacobian is a 4-vector, determined by simple differentiation:
\begin{equation}
\mathbf{J}_\mathrm{bin} = -\left( \sum \frac{\Delta_{\mathrm{data},i}}{\sigma_i^2}\nabla \Delta_{\mathrm{data},i} 
+ \frac{\Delta_\rho}{\sigma_\rho^2}\nabla \Delta_\rho \right).
\end{equation}


The Hessian is written as

\begin{equation}
\mathbf{H}_{lm} = -\left( \sum_i \frac{\left[\nabla_l\Delta_{\mathrm{data},i} \nabla_m\Delta_{\mathrm{data},i}\right] +\Delta_{\mathrm{data},i}\nabla^2_{lm} \Delta_{\mathrm{data},i}}{\sigma_i^2}
+\frac{\left[\nabla_l \Delta_\rho \nabla_m \Delta_\rho \right] + \Delta_\rho \nabla^2_{lm} \Delta_\rho}{\sigma_\rho^2} \right).\\
\end{equation}

The Jacobian and Hessian are thus merely constructed from the weighed sum of first and second derivatives of $\ln g$ and $\ln k$ (via $\Delta_\mathrm{data}$ and $\Delta_\rho$ respectively). We list the values of these terms in appendix \ref{mrp:app:cov_terms}.

%
%
%
%

\subsection{Method II: Sampled Data}
\label{mrp:app:jac_hess_po}

The most typical case will be that of sampled data, for which the general likelihood was presented as Eq. \ref{mrp:eq:basic_lnl}.
Since each of the functions $V$, $\phi\equiv g$ and $\rho_i$ may be dependent on $\thp$, it is not useful to specify the derivatives of the likelihood here -- often they will only be determined numerically, and as such it is simpler then to directly take the derivative of the likelihood itself. 

Nevertheless, it is perhaps helpful to specify the simplest special case, in which $V$ and $\rho$ are independent of $\thp$. 
In this case, the jacobian is 
\begin{equation}
	\mathbf{J}_l = -\int V g_l dm + \sum_{i=1}^{N_{\rm data}} \frac{\int V g_l \rho(x_i)dm}{\int V g\rho(x_i)dm},
\end{equation}
and the Hessian is 
\begin{equation}
\mathbf{H}_{lm} = -\int V g_{lm} dm + \sum_{i=1}^{N_{\rm data}}\frac{\int V g_{lm}\rho(x_i)dm}{\int Vg\rho(x_i)dm} - \frac{\int V g_l \rho(x_i)dm\int V g_m \rho(x_i)dm}{\left(\int V g \rho(x_i)dm\right)^2},
\end{equation}
where $g_{lm}$ is the (double) derivative of $g$ with respect to $\thp_l$ and $\thp_m$.



These equations are true given that uniform priors are used for all parameters. Insertion of a non-uniform prior requires the addition of the derivatives of the logarithm of the prior distribution. 

\subsubsection{Special Case: Power-Law Volume Function}
\label{mrp:app:special_case:pl}
The most complex special case that can be dealt with analytically is that in which the mass uncertainties are zero -- i.e. $\rho$ is a delta function -- and $V(m) = V_0 \kappa m$.
Unfortunately, the most common used uncertainty distribution for the masses is a log-normal distribution (i.e. a normal distribution in $m$), which renders the integral over $g\rho$ non-analytic.
The form of $V$ taken here is reasonably general -- it is a power-law, which is a reasonable form to expect physically, and to that we add an optional lower truncation at $\mmin$, in the case that $\kappa$ is is too low to ensure convergence in density. 

In this case, the form of $Vg$ is equivalent to $g$ with modified parameters, i.e. $V_0 \kappa m g(\hs,\alpha,\beta,\ln A) = g(\hs, \alpha+\kappa,\beta, \ln A+ \ln V_0 + \ln10 \kappa\hs)$, and we will denote the latter simply as $\tilde{g}$.
The Jacobian of this system is thus
\begin{equation}
	\mathbf{J}_l = -\tilde{q}_l + \sum_{i=1}^{N_{\rm data}} \frac{\tilde{g}^i_l}{\tilde{g}^i} \equiv -\tilde{q}_l + \sum_{i=1}^{N_{\rm data}} \ln \tilde{g}^i_l,
\end{equation}
and the Hessian is
\begin{equation}
\mathbf{H}_{lm} = -\tilde{q}_{lm} + \sum_{i=1}^{N_{\rm data}} \frac{\tilde{g}^i_{lm}}{\tilde{g}^i} - \frac{\tilde{g}^i_l \tilde{g}^i_m}{\tilde{g}_i^2} \equiv  -\tilde{q}_{lm} + \sum_{i=1}^{N_{\rm data}} \ln \tilde{g}^i_{lm}.
\end{equation}

The multivariate chain rule may be used to relate $\tilde{g}_l$ to $g_l$.

\subsection{Method III: Expected Data}
\label{mrp:app:expected}
In \S\ref{mrp:sec:expected_cov} we presented the expected covariance at the MLE for a sample drawn from a given model, and fit with the same model.
In this section, we present the general likelihood in this framework, and also some special cases in the context of the MRP.

The principle observation to make is that the sum in Eq. \ref{mrp:eq:basic_lnl} is over the samples, which must be drawn from some model distribution. 
The expected value of this distribution for any observed $\vec{x}'$ may be written $n'(\vec{x}')$, where the prime represents an observed quantity.
In the limit of infinitessimal bin-width, the sum becomes an integral over the observations:
\begin{equation}
	\label{eq:basis_lnl_expected}
	\ln \mathcal{L} = -\int V \phi dm + \int n'(\vec{x}') \ln \int V \phi \rho(\vec{x}') dm d\vec{x}'.
\end{equation}

There are two main special cases of interest, as well as their conjunction, in the context of the MRP.

\subsubsection{Special Case I: Power-Law Volume Function}
Here we revisit the special case of \S\ref{mrp:app:special_case:pl} to determine the expected likelihood.
Here, the likelihood is
\begin{equation}
 \ln \mathcal{L} = -\tilde{q} + \int n'(m) \ln \tilde{g}(m) dx,
\end{equation}
with expected jacobian and hessian given as
\begin{align}
	\mathbf{J}_l &=  -\tilde{q}_l + \int n'(m) \ln \tilde{g}_l dm \\
	\mathbf{H}_{lm} &= -\tilde{q}_{lm} + \int n'(m) \ln \tilde{g}_{lm} dm.
\end{align}

\subsubsection{Special Case II: Correct Data Model}
In this case, we assume that the model from which the data was actually drawn is the same as that being fit, such that the MLE of the parameters $\thp$ is equivalent to the true parameters of the data.
Thus we have 
\begin{equation}
	n'(x') = \int V' g' \rho'(\vec{x}') dm.
\end{equation}

The application of this formula to the likelihood, Eq. \ref{eq:basis_lnl_expected} is trivial. More interesting is its application to the Jacobian and Hessian at the point of the MLE. 
Here, we use the notation $n^\theta$ to refer to the total expected number of observations, equivalent to $\int V g dm $. 
The jacobian is
\begin{equation}
	\mathbf{J}_l = -n^\theta_l + \int n'(x') \frac{n_l(\vec{x}')}{n(\vec{x}')} d\vec{x}',
\end{equation}
where $n(x')$ is the model-dependent number density of observations at $x'$ (i.e. the same formula as $n'$ with primes removed). At the MLE, $n \equiv n'$, so that only $n_l(\vec{x}')$ remains within the integral. 
However, the expected total density of observed objects is the same as the total number of true objects (i.e. the integral of $\rho(x')$ is unity), so that the first and second terms cancel each other, leaving a jacobian of equivalently zero,a s we should expect.

Using a similar method for the hessian, we arrive at
\begin{equation}
	\label{eq:expected_cov_2}
	\mathbf{H}_{lm} = - \int \frac{n_l(\vec{x}'|\hat{\theta})n_m(\vec{x}'|\hat{\theta})}{n(\vec{x}'|\hat{\theta})} d\vec{x}',
\end{equation}
which is just Eq. \ref{mrp:eq:expected_hessian}. 
This equation is important, as it specifies a hessian only in terms of first-order derivatives. 
Even in cases where the derivative of $n(\vec{x}')$ must be attained numerically, this provides a significant performance and accuracy boost to the hessian calculation.
	
\subsubsection{Special Case III: Conjunction of Special Cases}
\label{mrp:app:special_case_3}

Here we take the conjunction of the previous two special cases, such that $n(\vec{x}') = \tilde{g}$. 
This case is very useful, as we can determine a completely analytic solution to both the likelihood and covariance at the MLE. 
The results can be used to quickly infer the amount of information in the raw MRP model at a given set of parameters, and also to determine optimal power-law slopes for the volume function.

The likelihood at any $\thp$, for data drawn from $\thp'$, is
\begin{equation}
	\ln \mathcal{L} = -\tilde{q} + \int \tilde{g}' \ln \tilde{g} dm.
\end{equation}

For the remainder of the derivation, we omit the tilde notation, since every variable is in the transformed variables.
To evaluate the integral, we can use integration by parts. We note that the indefinite integral
\begin{equation}
\int g dm = A \Hs \gamma(z,x) \equiv A()q_0 - q(m)),
\end{equation}
where $\gamma$ is the lower-incomplete gamma function, and use it to find
\begin{equation}
\int_{m_\mathrm{min}}^\infty g' \ln g dm = q'\ln g_{\mathrm{min}} + A'\Hs'\int_{m_\mathrm{min}}^\infty -\gamma(z',x') \frac{d \ln g}{dm}dm,
\end{equation}
where the subscript $\mathrm{min}$ refers to the fact that here $g$ is a function of the (optional) truncation mass $m_\mathrm{min}$, and we have
\begin{equation}
\frac{d \ln g}{dm} = \frac{\alpha+1}{m} - \frac{\beta x}{m}.
\end{equation}

Compiling everything, we have 
\begin{equation}
	\label{eq:lnl_expected_anl}
\ln \mathcal{L} = -q + q' \ln g_\mathrm{min} + A'\Hs' \left((\alpha+1)t' - u\right),
\end{equation}
where 
\begin{equation}
t' =  \int_{\mmin}^\infty \left(\Gamma(z',x')-\Gamma(z') \right) \frac{dm}{m}
\end{equation}
and 
\begin{equation}
u =  \int_{\mmin}^\infty \left(\Gamma(z',x')-\Gamma(z') \right) \frac{\beta x}{m} dm .
\end{equation}
For both, we use the substitution $dm = m dx'/(\beta' x')$, and for $u$ we also use integration by parts to finally yield
\begin{equation}
u = \left(\frac{\Hs'}{\Hs}\right)^\beta\left[\Gamma\left(z'+\frac{\beta}{\beta'},x'\right) - \frac{x'^{\beta/\beta'}q'}{\Hs'}\right]
\end{equation}
and 
\begin{equation}
\begin{split}
t' =  \frac{ \Gamma(z')}{\beta'}\biggl[&y'^{\alpha'+1}\Gamma(z') {}_2F_2\left( \begin{array}{cc} z' & z' \\ z'+1 & z'+1 \end{array},-x'\right)\\
&+\mathrm{Poly}\Gamma(z') - \beta'\ln y'\biggr],
\end{split}
\end{equation}
with ${}_pF_q$ the generalised hypergeometric function and $y = 10^{m-\hs}$.

To derive the hessian, it makes the most sense to begin with Eq. \ref{eq:lnl_expected_anl} rather than Eq. \ref{eq:expected_cov_2}. We find \begin{equation}
\label{mrp:eq:dlnLdXdY}
\mathbf{H}_{lm} = -q_{lm} + q' \ln g_{\mathrm{min},lm}  -  A'\Hs' u_{lm}.
\end{equation}

Expressions for the derivatives of $\ln g$, $q$ and $u$ can be found in \S\ref{mrp:app:cov_terms} (note that in this case, the derivatives of $\ln g$ have $m$ replaced with $m_{\mathrm min}$).

\subsection{Derivatives of \textit{g}, \textit{q}, \textit{k} and \textit{u}}
\label{mrp:app:cov_terms}
Several of the Jacobians and Hessians listed in this section require the derivatives of $\ln g$, $q$, $\ln k$ and $u$. Here we list these as coupled equations. Throughout the following we denote partial derivatives with respect to any parameter by a subscript, and in each case, the order of parameters in the jacobian/hessian is ($\hs,\alpha,\beta,\ln A$).

We begin with $\ln g$:
\begin{equation}
	\mathbf{J}_{\ln g} = \left\{\ln(10)(\beta x_m-\alpha), \ln y, \frac{1}{\beta}\left(1 - x_m\ln x_m\right), 1 \right\}
\end{equation}
\begin{equation}
\mathbf{H}_{\ln g} = \begin{pmatrix}
  -\ln^2 (10) \beta^2 x & -\ln 10 & \ln(10) x (1+\ln x) & 0 \\[2ex]
  . & 0 &  0 & 0 \\
  . & . & -\dfrac{1}{\beta^2}\left(1+x\ln^2 x\right) & 0 \\
  . & . & . & 0 
\end{pmatrix}
\end{equation}

For the remaining, it is helpful if we first define a couple of functions based on the Meijer-G function $G$:
\begin{equation}
    G1(z,x) = G^{3,0}_{2,3} \left(x\left| \begin{array}{ccc}
    1 & 1 &  \\
    0 & 0 & z
    \end{array}\right.\right)/\Gamma(z,x), \\
\end{equation}
\begin{equation}
    G2(z,x) = G^{4,0}_{3,4} \left(x\left| \begin{array}{cccc}
    1 & 1 & 1 & \\
    0 & 0 & 0 & z
    \end{array}\right.\right)/\Gamma(z,x),
\end{equation}
and
\begin{equation}
\bar{G}(z,x) = G1(z,x)\ln x + 2G2(z,x).
\end{equation}
 Furthermore we use the following commonly occurring relation:
 \begin{equation}
 \Phi = yg/\Gamma(z,x).
 \end{equation}
 
We note that everywhere in the derivatives of $q$, instances of $m$ are taken to rather be $m_{\mathrm min}$. Keeping this in mind, for the first derivatives of $\ln q$, we have 
\begin{equation}
    \mathbf{J}_q = q\left\{ \ln(10)(1+\Phi), \beta^{-1} (\ln x + G1), -z \ln q_\alpha - \frac{\ln y}{\beta}\Phi, 1  \right\} \\
\end{equation}
and for the double-derivatives
\begin{equation}
\mathbf{H}_q = \begin{pmatrix}
   \ln(10)\left(q_h + q\Phi \ln g_h\right) & \ln(10) \left(q_\alpha+ q\Phi \ln g_\alpha\right) & \ln(10)\left(q_\beta + q\Phi \ln g_\beta\right) & q_h \\[2ex]
   . & \beta^{-1} \left(-q_\alpha + q_\beta\ln x - q\dfrac{z\bar{G}}{\beta}\right) & \beta^{-1} \left(-q_\alpha + q_\beta\ln x - q\dfrac{z\bar{G}}{\beta}\right) & q_\alpha \\[2ex]
   . & . & q \dfrac{\Phi \ln y}{\beta} (\dfrac{1}{\beta}-\ln g_\beta)-z( q_{\alpha,\beta}-\dfrac{q_\alpha}{\beta}) & q_\beta \\[2ex]
   . & . & . & q
\end{pmatrix}.
\end{equation}


The derivatives of $\ln k$:
\begin{equation}
	\mathbf{J}^3_{\ln k} = \left\{ 2\ln(10), \mathrm{Poly}\Gamma(0,z_k)/\beta -z_k \ln k_\alpha \right\}
\end{equation}
and 
\begin{equation}
	\mathbf{H}^3_{\ln k} = \begin{pmatrix}
	0 & 0 & 0 \\
	. & \mathrm{Poly}\Gamma(1,z_k)/\beta^2  & -\left(z_k\ln k_{\alpha,\alpha} + \frac{\ln k_\alpha}{\beta}\right) \\
	. & . & \ln k_{\alpha,\alpha}z_k^2 + z_k\frac{\ln k_\alpha}{\beta}
	\end{pmatrix}.
\end{equation}

Finally, the derivatives of $u$:
\begin{equation}
\mathbf{J}^3_u = \left\{-\ln(10) \beta u, 0, \frac{u \ln x +  G1\cdot\Gamma}{\beta} \right\},\\
\end{equation}
and
\begin{equation}
\mathbf{H}^3_u = \begin{pmatrix}
-\ln(10)\beta u_h & 0 & -\ln(10)(u+\beta u_\beta)\\[2ex]
. & 0 & 0  \\[2ex] 
. & . & \dfrac{u_\beta \ln x + \bar{G}\cdot \Gamma/\beta}{\beta} \\[2ex] 
\end{pmatrix}
\end{equation}
where $\Gamma$, $G1$ and $\bar{G}$ are here functions of $(z+1,x)$. Note that these formulae for the partial derivatives of $u$ are only strictly correct at the solution (in general they involve $\thp'$ as well as $\thp$). The accompanying implementations calculate the more general form.

\label{mrp:lastpage}

\end{appendix}
\end{document}
\fi